\documentclass[acmlarge, authorversion, nonacm]{acmart}

\usepackage{gensymb}
\usepackage{amsmath}
\usepackage{subfig}
\usepackage{tabularx}
\usepackage{booktabs}

\usepackage[normalem]{ulem}

\AtBeginDocument{%
  }

\setcopyright{acmlicensed}
\acmJournal{IMWUT}
\acmYear{2024} \acmVolume{8} \acmNumber{4} \acmArticle{183} \acmMonth{12}\acmDOI{10.1145/3699752}

\begin{document}

\newcommand{\name}{ActSonic}{} 

\newcommand{\add}[1]{\textcolor{black}{{#1}}}
\newcommand{\remove}[1]{{\sout{ #1}}}


\title{\name{}: Recognizing Everyday Activities from Inaudible Acoustic Wave Around the Body}






\author{Saif Mahmud}
\affiliation{%
  \institution{Cornell University}
  \city{Ithaca, NY}
  \country{USA}}
\email{sm2446@cornell.edu}
\orcid{0000-0002-5283-0765}
\authornote{Corresponding Author}

\author{Vineet Parikh}
\affiliation{%
  \institution{Cornell University}
  \city{Ithaca, NY}
  \country{USA}}
\email{vap43@cornell.edu}
\orcid{0009-0000-8791-9340}

\author{Qikang Liang}
\affiliation{%
  \institution{Cornell University}
  \city{Ithaca, NY}
  \country{USA}}
\email{ql75@cornell.edu}
\orcid{0009-0001-9301-625X}

\author{Ke Li}
\affiliation{%
  \institution{Cornell University}
  \city{Ithaca, NY}
  \country{USA}}
\email{kl975@cornell.edu}
\orcid{0000-0002-4208-7904}

\author{Ruidong Zhang}
\affiliation{%
  \institution{Cornell University}
  \city{Ithaca, NY}
  \country{USA}}
\email{rz379@cornell.edu}
\orcid{0000-0001-8329-0522}

\author{Ashwin Ajit}
\affiliation{%
  \institution{Cornell University}
  \city{Ithaca, NY}
  \country{USA}}
\email{aa794@cornell.edu}
\orcid{0009-0009-0700-688X}

\author{Vipin Gunda}
\affiliation{%
  \institution{Cornell University}
  \city{Ithaca, NY}
  \country{USA}}
\email{vg245@cornell.edu}
\orcid{0009-0000-5500-2183}

\author{Devansh Agarwal}
\affiliation{%
  \institution{Cornell University}
  \city{Ithaca, NY}
  \country{USA}}
\email{da398@cornell.edu}
\orcid{0009-0005-1338-9275}

\author{Francois Guimbretiere}
\affiliation{%
  \institution{Cornell University}
  \city{Ithaca, NY}
  \country{USA}}
\email{francois@cs.cornell.edu}
\orcid{0000-0002-5510-6799}

\author{Cheng Zhang}
\affiliation{%
  \institution{Cornell University}
  \city{Ithaca, NY}
  \country{USA}}
\email{chengzhang@cornell.edu}
\orcid{0000-0002-5079-5927}

\renewcommand{\shortauthors}{Mahmud et al.}

\begin{abstract}


We present \name{}, an intelligent, low-power active acoustic sensing system integrated into eyeglasses that can recognize 27 different everyday activities (e.g., eating, drinking, toothbrushing) from inaudible acoustic waves around the body. It requires only a pair of miniature speakers and microphones mounted on each hinge of the eyeglasses to emit ultrasonic waves, creating an acoustic aura around the body. The acoustic signals are reflected based on the position and motion of various body parts, captured by the microphones, and analyzed by a customized self-supervised deep learning framework to infer the performed activities \add{on a remote device such as a mobile phone or cloud server}. \name{} was \add{evaluated} in user studies with 19 participants across 19 households \add{to track its efficacy in everyday activity recognition}. Without requiring any training data from new users (leave-one-participant-out evaluation), \name{} detected 27 activities, achieving an average F1-score of 86.6\% in fully unconstrained scenarios and 93.4\% in prompted settings at participants' homes.



\end{abstract}

\ccsdesc[500]{Human-centered computing~Ubiquitous and mobile computing systems and tools}

\acmArticleType{Research}
\keywords{Acoustic Sensing; Activity Recognition, Self-supervised Leaning}

\maketitle
\newif\iffvg
\fvgfalse

\section{INTRODUCTION}
\label{sec:intro}

\begin{figure*}[t!]
    \centering
    \includegraphics[width=\linewidth]{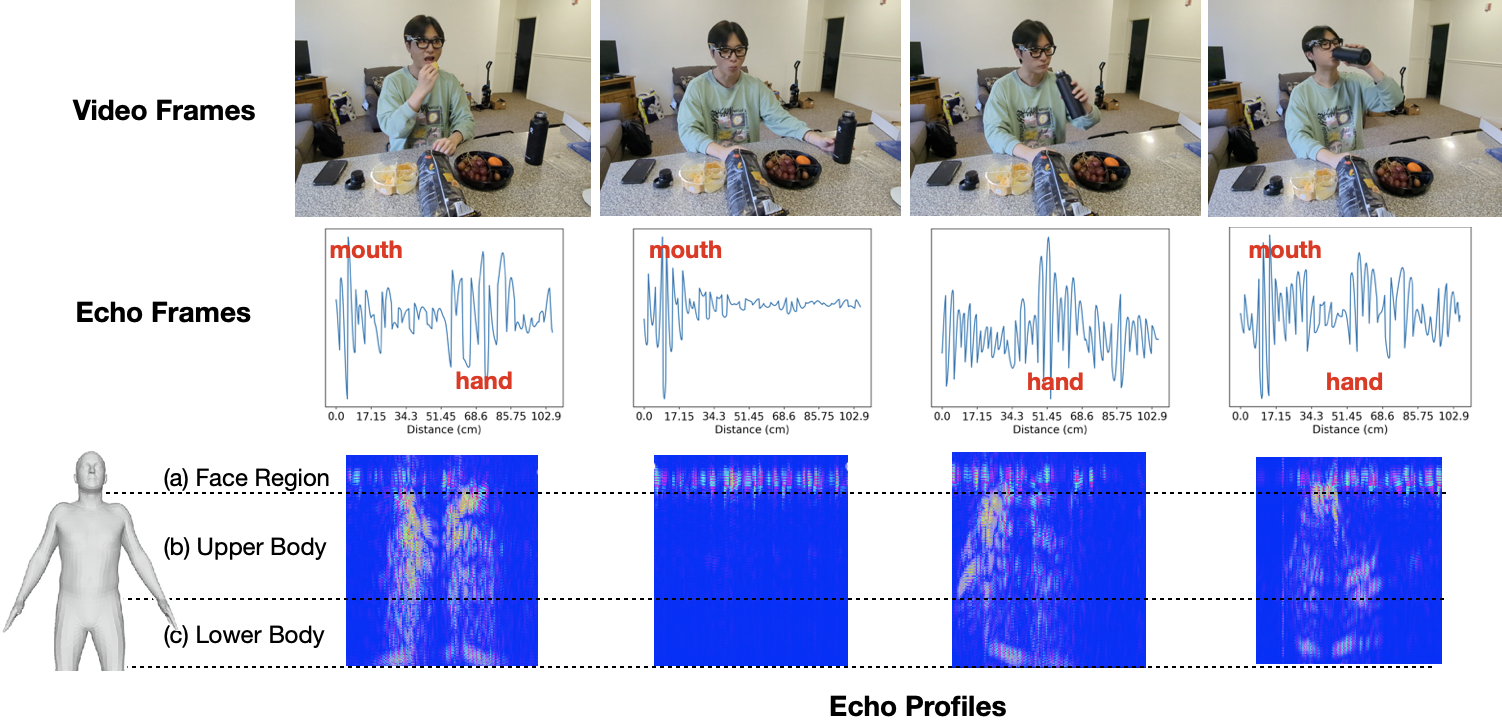}
    \caption{Overview of the active acoustic sensing principle of \name{}: The $x$-axis of the echo frames (in the 2nd row) represents the distance of echo reception. The corresponding video frames (in the first row) serve as activity references. The echo profile, created by stacking multiple echo frames, provides a spatiotemporal representation of the activity. These sliding windows with a duration of $2$ seconds of echo profiles (in the 3rd row) serve as inputs for the self-supervised learning algorithm.}
    \label{fig:echo-frames}
\end{figure*}


Despite smart glasses becoming popular and being worn daily, existing eyeglasses like many other commodity wearables, still lack the ability to understand the user's fine-grained everyday activities in real-world settings. Distinguishing human activities is challenging due to the substantial variance across different activities, each involving complex and subtle movements of multiple body parts (e.g., face, arms, hands). For example, eating requires the coordination of fingers and arms to deliver food to the mouth, as well as the movement of facial muscles for chewing.

Acquiring precise body pose data from various body parts is essential for accurately distinguishing different human activities. However, achieving this with everyday wearable devices (e.g., eyeglasses) has presented persistent challenges: wearable cameras~\cite{price2022unweavenet, liu2023diffusion} consume significant power, rapidly draining batteries. Furthermore, transmitting and processing video data on wearable devices is resource-intensive. Many sensing systems (Inertial Measurement Units (IMUs)~\cite{tong2020accelerometers, eatingtrak} or Electromyography (EMG)~\cite{zhang2017monitoring, zhang2018free}) only capture information around the instrumentation position, which is insufficient for activities involving multiple body parts (e.g., face and hands) unless multiple devices are worn on the body, which is less preferred by many users. Microphone-recorded passive sounds~\cite{iravantchi2021privacymic, mollyn2022samosa, laput2018ubicoustics} have shown promising performance in tracking human activities in the wild. However, it does not work effectively for activities lacking distinctive auditory cues. Therefore, achieving accurate recognition of each nuanced activity often necessitates a new set of customizations on wearables with carefully selected sensor combinations, a solution that may not be generalized well across different activities. Unlike the computer vision community, which primarily relies on cameras as the sole type of sensor to record data, facilitating the establishment of various benchmark datasets, researchers in the wearable and ubiquitous computing community face challenges in collecting data to track fine-grained human behaviors.

In this paper, we aim to answer the question: \textit{Is it possible to develop a simple sensing system on eyeglasses that is low-cost, minimally-obtrusive, energy-efficient, and privacy-aware, while capable of recognizing a wide range of everyday human activities? } A positive answer would immediately lower the barriers for researchers to study and track human activities of their interest.

To tackle this challenge, we utilized the empirical observation that many human activities involve both facial and upper-body movements. By using a single sensing system on eyeglasses to simultaneously capture high-quality facial movements and upper body poses, this information can potentially be processed by a customized deep learning framework to distinguish a wide range of human activities. However, designing such a wearable system that accurately recognizes various human activities with high resolution is highly challenging: 

\begin{itemize}
    \item The system must operate on low power to sustain continuous usage throughout the day.
    \item Privacy concerns of the user on wearing such a wearable system daily can be potentially alleviated with a carefully designed solution.
    \item The appearance of the wearable needs to be largely unchanged while wearing comfort is not compromised.
    \item The system must demonstrate robustness to accommodate various daily activities and distinguish them from other categories, including "Null", which refers to activities not included in the tracking set.
    \item Demonstrating the efficacy of the system with high temporal resolution requires recording and labeling ground truth data at the second level, a challenging and time-consuming task that many prior works overlooked.
\end{itemize}

Recent research has demonstrated the feasibility of incorporating active acoustic sensing into eyeglasses using commodity microphones and speakers. By analyzing the reflected acoustic waves from the eyes, face, or upper body, it becomes possible to infer gaze~\cite{li2024gazetrak}, facial expression~\cite{earface, li2024eyeecho}, or upper body pose~\cite{PoseSonic, SonicASL, fan2023apg} respectively. This development underscores the potential of leveraging a single sensing modality—active acoustic sensing on eyeglasses—to simultaneously capture both facial movements and upper body poses, which can be used for activity recognition. In this paper, we aim to answer the \textit{research question}:

\begin{itemize}
    \item  Can we utilize the inaudible active acoustic signals reflected around the human body to recognize a rich set of everyday human activities? 
\end{itemize}

Given the small size, relatively low power consumption, low cost, and ubiquity of microphones and speakers, such eyeglasses capable of tracking diverse human activities would significantly reduce the barriers to activity monitoring, a critical task within the ubiquitous computing community. Instead of relying on multiple devices for different activities, this all-in-one system could effectively track and differentiate between multiple activities simultaneously. However, designing such an active-acoustic sensing system and proving its efficacy for everyday activity recognition presents significant challenges:

\begin{itemize}
\item While active acoustic sensing has demonstrated effectiveness in recognizing facial expressions and upper body pose respectively, it remains unclear if the fused reflected acoustic signals provide adequate information to distinguish a rich set of fine-grained human activities.
\item Designing machine learning and signal processing algorithms that focus on temporal and structural differences in activities from these reflected acoustic signals, rather than individual body shapes, for usability to minimize user effort for training.
\item Robustness to various acoustic noises, such as multi-path acoustic reflections from the environmental setups (e.g., surrounding furniture) and potential overlap with the ultrasound band, is essential for handling different environmental conditions.
\item Capturing nuanced variations in activities, particularly in large motions where finer details are significant, is crucial. For example, recognizing activities with similar arm or facial movements (e.g., yawning vs. coughing) requires an understanding of the combination of movements and temporal patterns.
\item Robustness towards a large number of unknown or unlabeled activities in the "Null" category is necessary.
\end{itemize}

To address this research question, we developed \name{}, a self-supervised and low-power activity recognition system \add{placed on} eyeglasses form factor, utilizing active acoustic sensing. It is the first system to demonstrate the feasibility of recognizing 27 types of everyday activities without collecting training data from new users. Thanks to the low-power nature of acoustic sensors, \name{} \add{sensing system for data acquisition} can operate for over 21 hours with a battery capacity equivalent to that of Google Glass (570 mAh).

We developed \name{} by attaching a pair of miniature, low-power, off-the-shelf microphones and speakers to the hinges of glasses. The sensing system emits inaudible ultrasonic waves to create an \textit{acoustic aura} around the body. Based on the shape and position of various body parts, the acoustic signals are reflected with unique and fused patterns captured by the microphone. We designed a customized self-supervised deep learning framework\add{, running on a cloud server or Android mobile phone,} to interpret the reflected signals and their temporal patterns. These signals, presented with complex multipath echoes, include rich information about movements on both the face and upper body, allowing us to infer the performed activities.

\name{} was evaluated comprehensively in two studies in real-world settings. The first study was a semi-in-the-wild investigation involving 12 participants. In this study, each participant performed all 27 activities at their own homes (12 homes) in the presence of a researcher. To further validate the system's performance in completely uncontrolled real-world conditions, we conducted a second study with 7 participants. In this second study, participants were provided with the device at their homes to record their unconstrained daily activities alone without any intervention. These two studies resulted in the collection of 40 hours of activity data from 19 different households, which were labeled with ground truth at each second. The leave-one-participant-out evaluation showed that \name{} was able to distinguish these 27 activities at these 19 homes at each second without any training data from the user, with an average F1-score of 93.4\% in the first semi-in-the-wild user study and 86.6\% in the second in-the-wild study.

\name{} advance the field of eyeglasses-based activity recognition and lower the barrier of tracking everyday human activities in the following aspects: 

\begin{itemize}
    \item The first demonstration of creating and utilizing inaudible acoustic waves around the body \add{on eyeglasses form factor} for fine-grained everyday activity recognition. 
    \item Unlike many new data-driven activity recognition systems on wearables, \name{} achieved promising performance without requiring training data from a new user or a new environment, which makes collecting human activity data much easier. 
    \item Microphones and speakers are readily available, cost-effective to set up, and can sustain prolonged operation on a small battery attached to glasses, facilitating easy replication of our system by other researchers.  
    \item Compared to many previous works relying on multiple sensors or instrumentations, \name{} accurately recognizes 27 activities with a single \add{sensing module} at a high time resolution of one second. 
    
\end{itemize}

\section{RELATED WORK}
\label{sec:rel-work}

A large and growing body of literature on activity recognition has investigated various wearable and non-wearable sensing systems, including IMUs, cameras, microphones, water pressure, and powerline sensors~\cite{froehlich2009hydrosense, gupta2010electrisense}, as well as multimodal sensor fusions. In this section, we provide an overview of related work focused on IMU, camera, and acoustic sensing-based activity recognition, and position the contribution of \name{} within this landscape.

\subsection{IMU-based Human Activity Recognition}
Inertial Measurement Units (IMUs) in commodity smartwatches, phones, and other wearables have garnered considerable interest in detecting human activities over a significant period. These inertial sensors include accelerometers, gyroscopes, magnetometers, etc. Early research into IMU-based activity recognition relied on hand-crafted features~\cite{kwon2018adding, qian2019novel, bao2004activity, haresamudram2019role} generated from sensor readings. These methods~\cite{10.1145/1964897.1964918} were initially limited to recognizing coarse human locomotion activities such as walking, running, sitting, etc. With the advent of end-to-end deep learning methods~\cite{guan2017ensembles, hammerla2016deep, ordonez2016deep, murahari2018attention, ecai-har} to extract feature representations from time-domain sensor readings, IMU-based systems demonstrated an extended capability to recognize more fine-grained actions such as apparatus usage~\cite{gierad-smartwatch}, body gestures~\cite{gierad-biosense}, etc. These deep learning architectures, incorporating convolutions, recurrence, and transformers~\cite{ma2019attnsense, yao2019sadeepsense, kdd-har}, led to the detection of activities with high fidelity and low error rates. Self-supervision for IMU-based activity recognition is particularly effective in scenarios with small labeled datasets and exhibits significant performance improvement through the utilization of large unlabeled datasets. The pre-training tasks for these self-supervised models are designed as masked window reconstruction~\cite{haresamudram2020masked}, signal correspondence learning~\cite{saeed2020federated}, and contrastive predictive coding~\cite{harish-cpc-har} of temporal signals. Despite the success of IMU-based systems in detecting certain fine-grained activities, their capacity is limited due to the low spatial resolution~\cite{tong2020accelerometers} of the sensing modalities. Therefore, capturing a wide range of activities with a single placement of IMU remains challenging for wearables.

\subsection{Vision-based and Multimodal Human Activity Recognition}
Computer vision-based approaches incorporate cameras as egocentric wearables~\cite{price2022unweavenet, ohkawa2023assemblyhands,bodytrak,10.1145/3581641.3584063,10.1145/3379337.3415879,lim2023c} or systems installed in an environment~\cite{duan2022revisiting}. In vision-based action segmentation approaches~\cite{huang2016connectionist, lea2017temporal, farha2019ms, kukleva2019unsupervised, sarfraz2021temporally}, they are tasked with assigning activity labels to each frame of the video. These vision-based approaches adopt weakly supervised~\cite{kim2022detector} or unsupervised~\cite{kukleva2019unsupervised, sarfraz2021temporally} or self-supervised~\cite{price2022unweavenet} modeling techniques to detect activities by learning temporal embedding of the video frames. On the other hand, multimodal approaches~\cite{mm-fit, laser-eye-act, nadir-multimodal, attend-discriminate-har, act-forecast} utilize a fusion of sensing modalities to recognize human activities. These multimodal approaches obtain information from different combinations of IMUs, cameras, and microphones to recognize context-aware daily living activities~\cite{radu-multimodal, laput2017synthetic}, body or finger gestures~\cite{ward2006activity, lukowicz2004recognizing, zhang2017fingersound, sun2021teethtap,sun2021thumbtrak, fingerping}. Although vision or multimodal sensing-based activity recognition approaches demonstrate promising performance, they pose the challenge of privacy breaches and high power consumption.

\subsection{Acoustic Sensing-based Human Activity Recognition}
The aforementioned vision-based and multimodal sensing approaches for detecting human activities offer higher spatiotemporal resolution, leading to lower recognition errors. However, the usage of multiple sensors raises concerns about increased power consumption and privacy invasion. Recently, researchers have leveraged the pervasiveness of microphones in off-the-shelf commodity devices to recognize human activities~\cite{wu2020automated, liang2019audio, laput2018ubicoustics, lane2015deepear}. These acoustic sensing systems utilize passively sensed audio within the audible frequency range ($20$ Hz to $16$ KHz). Although passively sensed environmental sound provides discriminative information required to infer certain activities that generate sound, it raises serious privacy concerns since it may record personal conversations. In response to this, \add{existing works~\cite{mollyn2022samosa, chen2008audio, audio-har-mobilehci20}} adopt subsampling and other preprocessing techniques to make the audio unintelligible before activity recognition. Additionally, passively sensed audio from inaudible frequency ranges (infrasonic and ultrasonic) has been utilized to recognize daily activities~\cite{kijima2018multiple, iravantchi2021privacymic, iravantchi2019interferi}. While these techniques offer better preservation of user privacy, activity recognition on these systems relies on the assumption that activities will generate environmental sound that can be modeled by the system. However, many activities in daily living involve body-limb movement and do not necessarily generate distinctive sounds. Active acoustic sensing-based approaches emit high-frequency sound waves and leverage the reflected sounds to capture fine-grained \add{facial movements~\cite{eario, echospeech, 10.1145/3594738.3611358, li2024eyeecho}, hand poses~\cite{lee2024echowrist, fingerio, fingerping, active-acoustic-hand-gesture}, sign language gesture~\cite{SonicASL, TransASL}, body pose~\cite{PoseSonic}, activities~\cite{LASense}, gaze~\cite{li2024gazetrak}, vital signs~\cite{c-fmcw, nandakumar2015contactless}, fitness monitoring~\cite{HearFit}, and robot manipulation~\cite{robot_manipulation_01, robot_manipulation_02}.}

In \name{}, we model daily living activities based on movements in different parts of the human body. To achieve this, we utilize an active acoustic sensing platform placed on eyeglasses to capture these fine-grained movements and subsequently recognize activities.

\section{DESIGN AND IMPLEMENTATION OF SENSING SYSTEM}
\label{sec:sensing-mechanism}

Our goal is to develop a wearable activity recognition platform integrated into glasses, enabling the identification of a wide range of everyday activities based on the positions and movements of different body parts, as captured by active acoustic sensing technology. While \add{prior research~\cite{PoseSonic, eario, echospeech, eyeEcho, SonicASL, fingerio}} has underscored the efficacy of active acoustic sensing in detecting facial changes and upper body poses, no existing system has seamlessly integrated this sensing modality for the comprehensive task of classifying diverse daily activities in real-world scenarios. Given that most daily activities entail movements across various body parts, particularly the upper body, head, and face, our primary focus with \name{} is to capture and differentiate these activities based on the position and movement of these body areas, as detected by our sensing system. \name{} accomplishes this with a singular \add{sensor placement}, facilitating concurrent information capture. This section provides an overview of the active acoustic sensing setup, the process of feature extraction for activity recognition, and the hardware implementation, with a particular emphasis on the wearable form factor of the system.

\begin{figure}[h]
    \centering
    \includegraphics[width=0.8\linewidth]{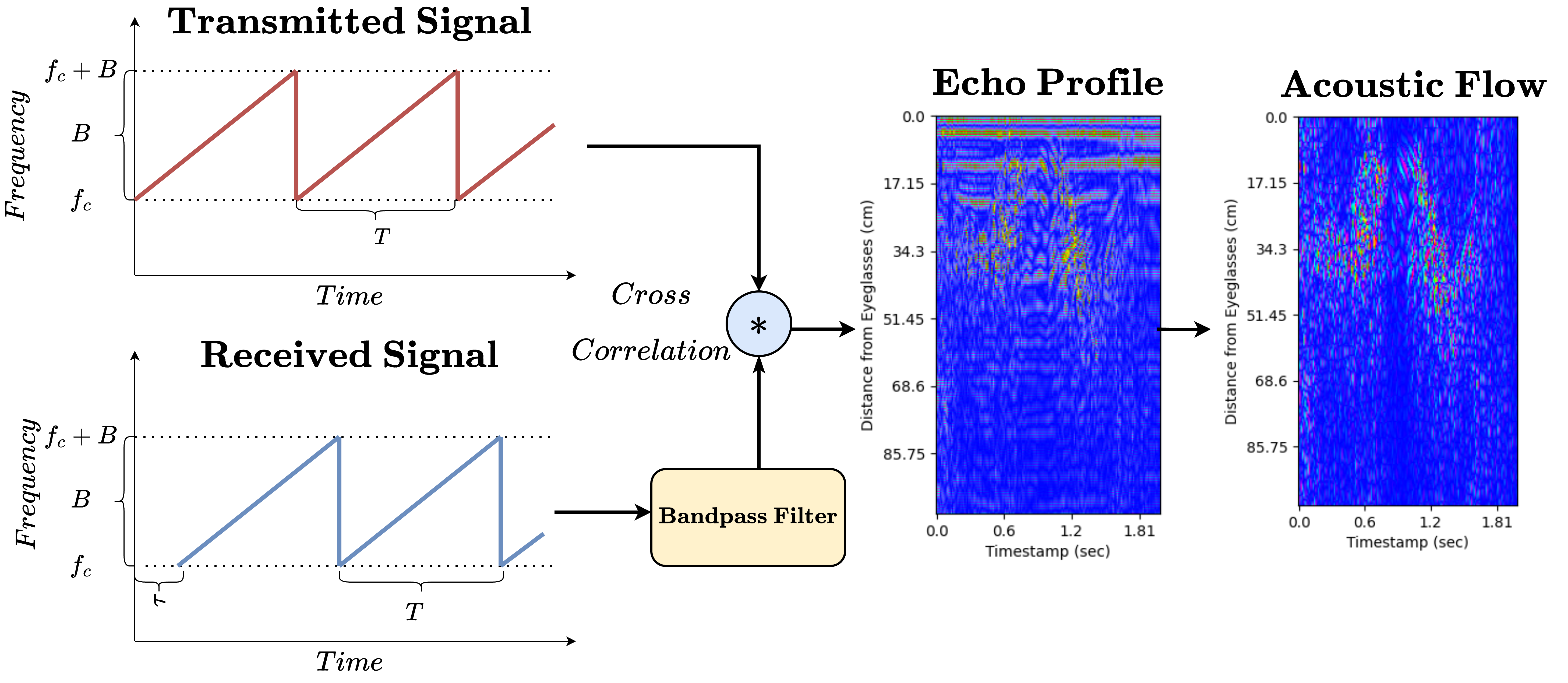}
    \caption{Overview of echo profile and acoustic flow calculation. For the echo profile, we cross-correlate the transmitted signal with a bandpass filter applied over the received signal (to ensure only specific frequencies are returned). This allows us to capture the direct echo profile, and we can calculate acoustic flow by taking the difference between two consecutive echo profiles.}
    \label{fig:fmcw}
\end{figure}

\subsection{Configuration of Active Acoustic Signal}
\label{sec:signal-config}
The active acoustic sensing system in \name{} incorporates two pairs of ultrasonic transmitters and receivers on the eyeglass hinges. Utilizing Cross-correlation-based Frequency Modulated Continuous Wave (C-FMCW)~\cite{c-fmcw} chirps, \name{} emits ultrasonic signals with linearly modulated frequencies ranging from $18.0$ to $21.0$ kHz and $21.5$ to $24.5$ kHz \add{simultaneously} for the left and right transmitters, respectively, each with a bandwidth of $B = 3$ kHz.

For the C-FMCW~\cite{c-fmcw} technique employed by \name{}, the theoretical range resolution $R$ can be derived as follows:

\begin{equation}
R = \frac{C \cdot Lag}{2 F_s} - vt
\label{eqn:fmcw-resolution-01}
\end{equation}

where $C$ is the speed of sound ($C = 343$ m/s in dry air at $20\degree$C), $F_s$ is the sampling frequency of the microphone or receiver ($F_s = 50$ kHz in the case of \name{}), $v$ is the velocity of the tracked body part, $t$ is the elapsed time from the start of the modulation period, and $Lag$ is the number of samples shifted between the transmitted and received acoustic signals. With $N = 600$ samples transmitted for each chirp, \name{}'s sensing system has a sweep period of $T = \frac{600}{50000} = 0.012$ seconds $= 12$ milliseconds. Thus, the value of $t$ ranges from $0 \le t \le 0.012$ s, and the value of $Lag$ in Eq.~\ref{eqn:fmcw-resolution-01} ranges from $0 \le Lag \le 600$.

According to the rationale provided in~\cite{c-fmcw}, we can set $v = 0$ in Eq.~\ref{eqn:fmcw-resolution-01} in the case where the tracked body part is static or moving slowly with respect to the speed of sound. The velocity of extremely fast arm movements, such as punching, has been measured at approximately $8.9$ to $11.5$ m/s~\cite{kimm2015hand}, while the duration of the fastest facial expression in humans is measured at 100 ms~\cite{yan2013fast}. The speed of these extreme body motions ($v$) is significantly smaller compared to the speed of sound ($C$). Therefore, based on the equation above, the impact of the sensing resolution ($vt$) caused by body movements is minimal. As a result, the theoretical range resolution $\delta R$ of the C-FMCW signal deployed in \name{} can be simplified to:

\begin{equation}
\delta R = \frac{C \cdot \delta Lag}{2 F_s}
\label{eqn:fmcw-resolution-02}
\end{equation}

Since cross-correlation is computed for each sample in the case of C-FMCW~\cite{c-fmcw}, we can set $\delta Lag = 1$ for the theoretical range resolution calculation in Eq.~\ref{eqn:fmcw-resolution-02}. Therefore, we can substitute the values of variables in Eq.~\ref{eqn:fmcw-resolution-02} and compute the resolution of the sensing system of \name{} as follows:

\begin{equation}
\delta R = \frac{343 \text{ m/s}}{2 \cdot 50000} = 0.343 \text{ cm} = 3.43 \text{ mm}
\label{eqn:fmcw-res-val}
\end{equation}

The maximum sensing range of \name{} can be computed as $R_{\text{max}} = N \cdot \delta R = 600 \cdot 0.343 \text{ cm} = 205.8 \text{ cm} \approx 2 \text{ m}$. This combination of high sensing resolution ($0.343$ cm) and extensive sensing range enables us to capture both subtle skin deformations on the face and monitor the pose and coarse movement of the upper body region effectively. Additionally, it is important to note that the sensing resolution mentioned above is calculated for range-finding applications, as demonstrated in~\cite{c-fmcw}, which differs from our approach. This calculation is provided as a reference to facilitate readers' understanding of the sensing principle underlying our proposed system.

\subsection{Computation of Echo Profile and Acoustic Flow}
\label{sec:echo-profile}


Active acoustic sensing mechanism in \name{} measures the round-trip delay between emitted and reflected ultrasonic waves to detect human body movements. To capture this delay ($\tau$), we utilize cross-correlation~\cite{c-fmcw} on transmitted and received signals. Figure~\ref{fig:fmcw} demonstrates the application of bandpass filters matching the transmitted frequency ranges ($18-21$ KHz and $21.5-24.5$ KHz) on received signals. This filtering eliminates audible frequencies, ensuring user privacy and eliminating environmental acoustic noise before cross-correlation computation.

The cross-correlation matches the sweep period of the emitted FMCW ultrasonic wave. It generates an \textit{echo frame}, represented as a $(600 \times 1)$ column vector in \name{}. Stacking these frames creates the \textit{echo profile}~\cite{fingerio, c-fmcw, eario, echospeech}, where brighter pixels indicate strong reflections at specific distances. With two transmitter-receiver pairs, \name{} accounts for four transmission paths. To elaborate, if we consider the left and right transmitter-receiver pair as $(T_{left}, R_{left})$ and $(T_{right}, R_{right})$ respectively, then the paths will be $T_{left} \rightarrow R_{left}$, $T_{left} \rightarrow R_{right}$, $T_{right} \rightarrow R_{left}$, and $T_{right} \rightarrow R_{right}$. In \name{}, we stack the outputs of cross-correlation of these four paths as four channels of the echo profile. Note that the computation of one channel in the echo profile is shown in Figure~\ref{fig:fmcw}.

Acoustic flow, also known as the differential echo profile, is derived by computing the derivative of distance from the eyeglasses (echo profile's $y$-axis) with respect to time ($x$-axis). This is achieved by calculating the absolute difference between consecutive echo frames. Acoustic flow effectively eliminates reflections from stationary objects, enabling precise detection of human body movements. Moreover, it mitigates the effects of eyeglasses remounting, ensuring a resilient measurement of body motion across sessions. The $y$-axis of the echo profile (from top to bottom) indicates the distance from the eyeglasses, while the $x$-axis represents the temporal axis. The sliding window employed in \name{} covers a sensing range of $300$ pixels, roughly equivalent to $1$ meter, extending up to the user's knees. This parameter of $300$ pixels has been fine-tuned to optimize activity recognition.

\subsection{Hardware Implementation and Wearable Form Factor}
\label{sec:hardware}


\begin{figure}[h]
    \centering
     \includegraphics[width=0.8\linewidth]{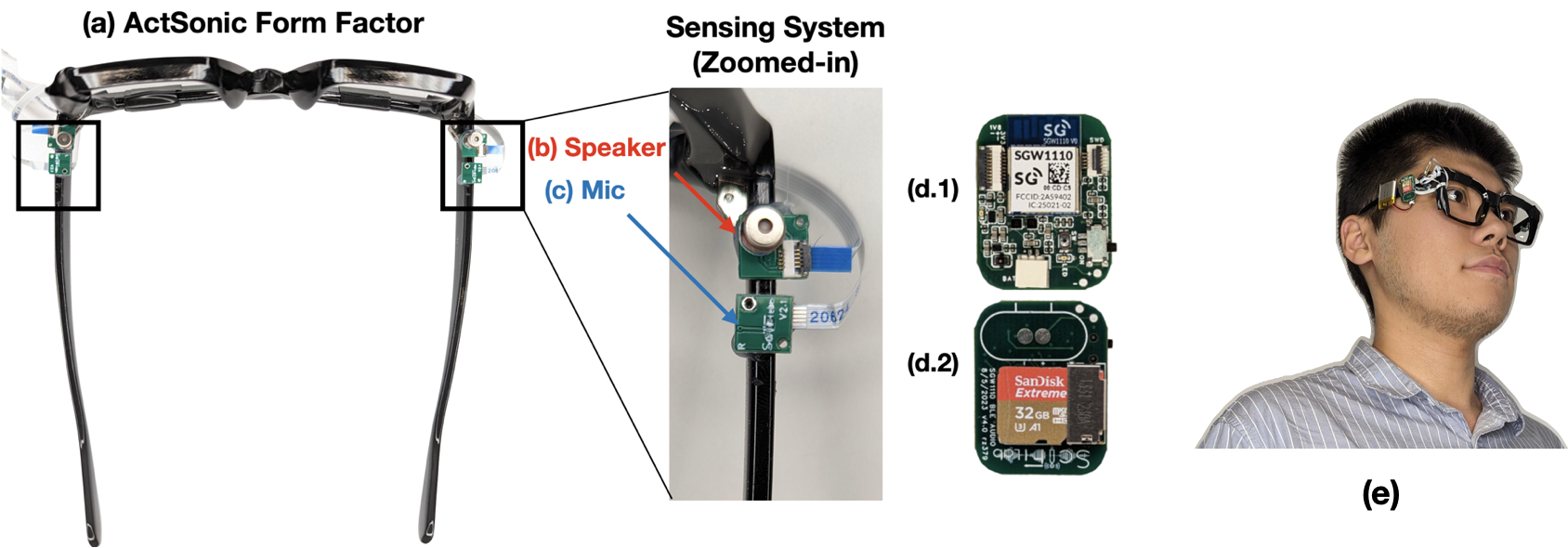}
     \caption{Hardware of \name{}: \textbf{(a)} Eyeglasses form factor,
     \textbf{(b)} Transmitter or speaker, \textbf{(c)} Receiver or microphone (dimension of the sensor board of (b) and (c) is $9 mm \times 9 mm$), (d) Front (d.1) and back (d.2) of customized PCB board (dimension $18 mm \times 23 mm$) with low-power nRF52840 micro-controller, (e) User wearing \name{} eyeglasses form factor}
     \label{fig:hardware}
 \end{figure}

We assembled the active acoustic sensing system for \name{} using two OWR-05049T-38D speakers and two ICS-43434 microphones~\cite{mic-i2s}, following a design similar to that shown in~\cite{eario}. Managed by a Teensy 4.1 microcontroller~\cite{teensy41}, the setup oversaw FMCW signal transmission and reception. To connect the speakers, microphones, and microcontroller, we developed a custom PCB housing two MAX98357A audio amplifier chips~\cite{amplifier}. Utilizing the Inter-IC Sound (I2S) interface, \name{}'s hardware components communicate, with received signals stored on an SD card via the micro-SD interface on the microcontroller.

Positioned symmetrically on a standard pair of glasses, the \name{} sensor system optimally captures nuanced body movements from various angles, facing perpendicularly downwards towards the body. After several design iterations, this orientation proved most effective for sound wave propagation. Connected via Flexible Printed Circuit (FPC) cables, the microcontroller, along with a Li-Po battery, is affixed to one leg of the glasses, interfacing with the speakers and microphones.

Initially, our prototype utilized the Teensy 4.1~\cite{teensy41} microcontroller, powered by an ARM M7 core, which exhibited higher power consumption due to its characteristics. \add{This Teensy 4.1~\cite{teensy41}-based prototype was used to collect data in the user studies intended to evaluate the system.} To highlight the power efficiency of our acoustic sensing, we designed a low-power variant featuring an nRF52840 microcontroller~\cite{nRF52840}, based on a low-power ARM M4 core. This variant includes two MAX98357A audio amplifiers, similar to the original setup, alongside power management modules and the SGW1110~\cite{ble-module} module. A 32 GB SanDisk Extreme microSD card~\cite{sd-card} handles storage, with optimized firmware minimizing SD card accesses for swift operations.

\section{DEEP LEARNING FRAMEWORK}
\label{sec:dl-model}

\begin{figure}[h]
    \centering
    \includegraphics[width=0.7\linewidth]{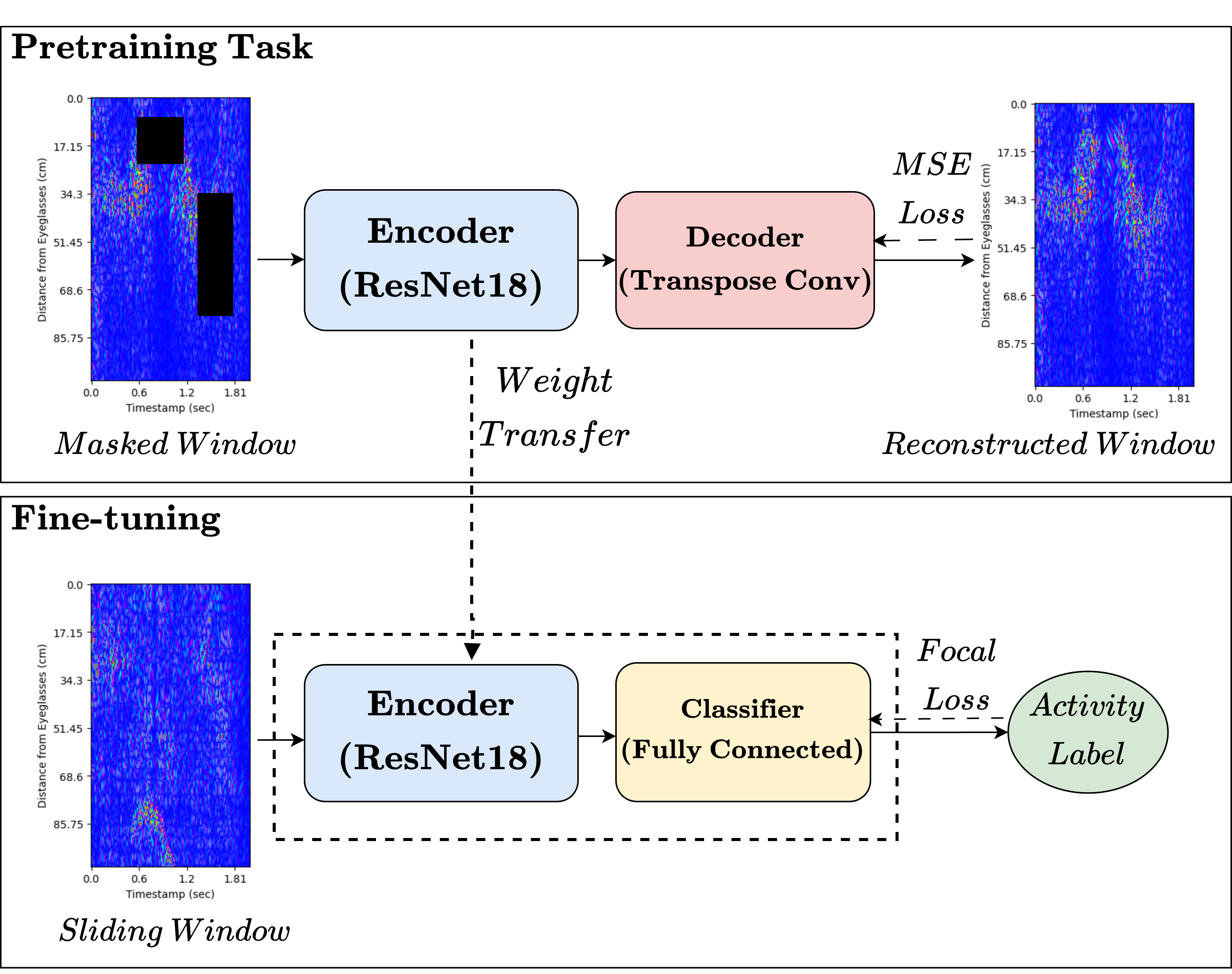}
    \caption{Deep learning model architecture for \name{}. Within the self-supervised \textbf{pretraining} stage, we mask out specific sections of the input echo profile and train an encoder-decoder architecture to reconstruct the input echo profile (given a lightweight decoder) supervised by an MSE loss. We then \textbf{fine-tune} the trained encoder from this step along with a lightweight classifier on the labeled dataset.}
    \label{fig:ml-model}
\end{figure}

To estimate human activities from acoustic sensing data in \name{}, we design a self-supervised deep learning framework which is illustrated in Figure~\ref{fig:ml-model}. This framework leverages the unlabelled acoustic data to create a pre-trained encoder. Then, we use this encoder to create an activity recognition pipeline.

\subsection{Self-supervised Learning Pipeline}
\label{sec:self-supervised}

Self-supervised learning is a form of supervised learning where the model predicts a subset of unlabelled data from the rest. This learning pipeline of \name{} consists of two steps: \textbf{pretraining} encoder to learn the representation of unlabelled data, and \textbf{fine-tuning} the pre-trained encoder weights for the target task with labeled data.

\subsubsection{Pretraining Task}
\label{subsec:pre-training}
The pre-training approach to learning representation from the unlabelled data is to perturb the sliding window of acoustic flow (described in~\ref{sec:echo-profile}) with binary mask and reconstruct the original window using the autoencoder depicted in the pre-training task segment of Figure~\ref{fig:ml-model}. Here, the binary mask is constructed in a way such that $m\%$ of each channel of the sliding window is set to $0$. In the case of \name{}, the numerical value $m$ of this mask percentage is randomly chosen from the range $15-20\%$, and the number of patches is randomly chosen from the range $1$ to $4$. As illustrated in Figure~\ref{fig:ml-model}, the masked window goes through a ResNet18~\cite{resnet} encoder. The embedded representation from the encoder is then fed through a decoder network, essentially a transpose convolution network. This customized transpose convolution, or deconvolution, network takes feature maps from the ResNet18 as input and generates a three-dimensional matrix of the shape of the input sliding window. We calculate the Mean Squared Error (MSE) between each pixel in the reconstructed and original sliding window as the loss function for this autoencoder architecture.

\subsubsection{Fine-tuning}
The aforementioned ResNet18 encoder of the pre-training pipeline (detailed in Subsec.~\ref{subsec:pre-training}) learns the representation of active acoustic data via self-supervision. This ResNet18 encoder with learned weights serves as the feature extraction pipeline in the fine-tuning phase. We design an activity recognition architecture (depicted in the fine-tuning segment of Figure~\ref{fig:ml-model}) comprising of pre-trained ResNet18 encoder followed by a fully connected classifier layer. We apply average pooling on the spatial axis of the feature map extracted by the encoder and feed it to the fully connected layer. The fully connected classifier network is a feedforward neural network with batch normalization~\cite{batch-norm}, Leaky ReLU activation~\cite{xu2015empirical}, and dropout~\cite{srivastava2014dropout} in between. We set the number of neurons in the last layer equal to the number of activity classes and perform a softmax operation to output a probability distribution.

The activity recognition model in the fine-tuning phase is trained using acoustic flow sliding windows as input and activity class labels as the target. To optimize the training process, we employ focal loss~\cite{lin2017focal} as the objective function. Focal loss is a modification of the standard cross-entropy loss, designed to emphasize learning from hard examples. It dynamically scales the loss function based on the confidence of the correct class prediction, with a decay factor that decreases as the confidence increases. In binary classification scenarios, where $p_t$ represents the predicted class probability, the standard cross-entropy loss $CE(p_t)$ is defined as $- \log(p_t)$ when $p_t = p$ for the positive class and $- \log(1 - p_t)$ otherwise. Focal loss adds a scaling factor $(1 - p_t)^\gamma$ to this standard cross-entropy loss, with $\gamma$ being a hyperparameter set to $0.5$ for \name{}. This modification ensures that the loss function assigns lower values to well-classified examples ($p_t > 0.5$) and focuses more on misclassified examples. This adaptation is particularly effective for \name{} due to the imbalanced distribution of activity labels in the dataset and the similarity in body motion patterns observed in the echo profile for some activities.

\subsection{Training and Implementation}
The self-supervised activity recognition model of \name{} processes overlapping sliding windows of the acoustic flow as input, with the shape of the sliding window being a hyperparameter. We conduct an iterative process to determine the optimal sliding window duration, ranging from $0.30$ seconds to $5.00$ seconds with a hop size of $0.10$ seconds. Performance evaluation on the validation set helps us fine-tune this parameter, resulting in an optimal duration of $2.00$ seconds with a $50$\% overlap. The shape of the input sliding window is defined as (num\_channels $\times$ num\_features $\times$ num\_samples) = ($4 \times 295 \times 166$). Here, num\_features represents the number of pixels from the echo profile, calibrated to $295$, covering a sensing range of approximately 1 meter (precisely $101.185$ cm), sufficient to capture upper body poses. Considering the sampling rate of \name{} at $50$ KHz and the number of samples in one sweep period at $600$ (details in Sec.~\ref{sec:sensing-mechanism}), a one-second sliding window contains approximately $\lfloor \frac{50000}{600} \rfloor = 83$ samples. Consequently, the num\_samples for a $2.00$-second sliding window of \name{} is set to 166.

The dropout probability of the feedforward classification layer in the fine-tuning phase is configured to 0.2. Both the pre-trained and fine-tuning models are trained for 100 and 50 epochs, respectively, using a batch size of 64. We employ the Adam optimizer \cite{kingma2014adam} and incorporate a cosine annealing learning rate scheduler with an initial learning rate of $10^{-3}$. The self-supervised model, including both the pre-training and fine-tuning networks, is implemented using the PyTorch and PyTorch Lightning frameworks and trained on GeForce RTX 2080 Ti GPUs.

\subsection{Evaluation Metric}
We use the Macro F1-score as our evaluation metric for \name{}'s activity recognition performance. If $C$ is the set of activity classes such that classes are indexed as $0, \dots, (C - 1)$ and $|C|$ is the cardinality of this set, the evaluation metric is defined as:
\begin{equation}
    Macro\ F1 = \frac{1}{|C|} \cdot \sum_{i = 0}^{C - 1} \frac{2 \cdot precision_i \cdot recall_i}{precision_i + recall_i}
\end{equation}

Where $precision_i$ and $recall_i$ are the numerical values of precision or positive predictive value and recall or sensitivity of $i$-th class respectively. 
\section{USER STUDY}
\label{sec:user-study}

In this section, we present a comprehensive overview of the user studies conducted to assess the performance of \name{}. The objective of these studies is to evaluate the activity recognition pipeline under naturalistic conditions. To achieve this goal, we devised a diverse set of everyday activities to monitor throughout the study, recruited participants, and carried out both a semi-in-the-wild user study and a more extended unconstrained study, both conducted in participants' homes in a naturalistic environment.

\subsection{Design of Activity Set}
\label{sec:activity-set}

To establish a set of activities, we conducted a pilot feasibility study encompassing over 50 activities of daily living with 5 users from our research team. Drawing on relevant prior studies~\cite{mollyn2022samosa, laput2018ubicoustics, iravantchi2021privacymic, edemekong2019activities} and insights from the pilot study, we selected 27 activities of daily living to be incorporated into \name{}'s tracking set. Additionally, we introduced a \textit{Null} label for activities not part of the tracking set. The primary criterion for selecting everyday activities was their involvement of movements across different body parts, aligning with \name{}'s reliance on tracking body motion. The activities in the tracking set are categorized into three segments based on their typical indoor locations:
\begin{itemize}
    \item \textbf{Bathroom (6 activities): } Rinse Mouth, Brushing Teeth, Flossing, Brushing Hair, Flush Toilet, Opening Door
    \item \textbf{Kitchen and Dining Area (8 activities) :} Washing Hands, Eating, Drinking, Pickup / Putdown, Pouring, Chopping, Wiping Surface, Stirring
    \item \textbf{Bedroom and Living Area (13 activities):} Stationary, Walking, Sitting, Coughing, Yawning, Talking, Putting on Outerwear, Vacuum, Throwing, Stretching, Using Phone/Tab, Squat, Reading Book
\end{itemize}

\subsection{Participants and Study Schedule}

\begin{figure}[h]
    \centering
    \includegraphics[width=0.75\linewidth]{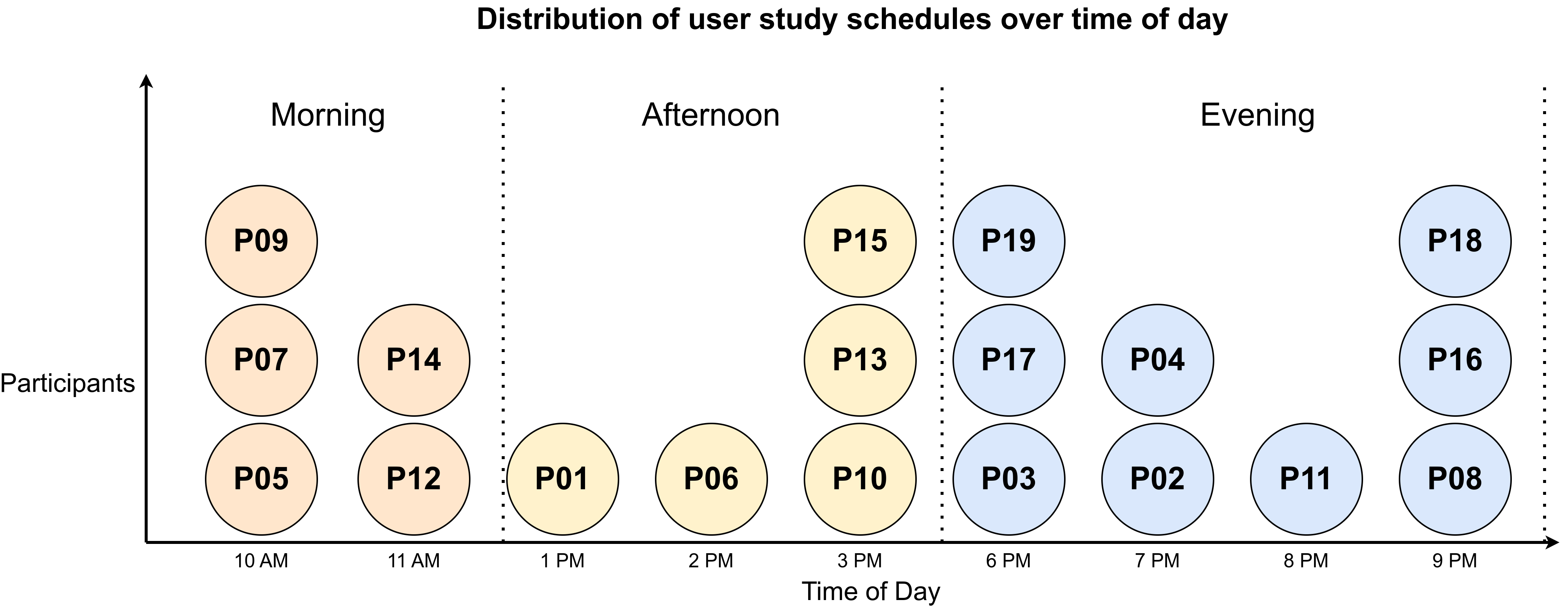}
    \caption{Distribution of participant schedules for the user study over time, where $x$-axis represents time and $y$-axis represents participant count. We split the participants into three general groups ("morning" as 7 am - 12 pm, "afternoon" as 12 pm - 6 pm, and "evening" as 6 pm - 11 pm) and ensure that we get a mixture of data across different times of day}
    \label{fig:usr-time-dist}
\end{figure}

The \name{} user studies received approval from the Institutional Review Board for Human Participant Research (IRB) at our organization. We enlisted 12 participants for a semi-in-the-wild study and 7 different participants for an unconstrained user study. Among the total 19 participants, with an average age of $(24.737 \pm 3.445)$ years, ranging from 21 to 31 years, 5 identified as female, and 14 identified as male. Per IRB guidelines, each study lasted no longer than 2 hours (120 minutes), and participants received \$30 USD as compensation for their time. Post-study, we gathered basic demographic and physical information (e.g., height, weight, gender), along with general feedback on the \name{} wearable device via an IRB-approved questionnaire.

The user studies took place in participants' homes, where they utilized their own tools or appliances as needed for activities. An exception was made for a few participants who were provided with dental floss for the flossing activity. A trained experimenter from our research team visited participants' addresses equipped with the necessary data collection apparatus to conduct the study.

Based on insights from a pilot feasibility study, human activity patterns vary throughout the day. For instance, activities such as tooth brushing are more likely to occur in the morning or after dinner. Accordingly, we scheduled user studies to capture activity data across different parts of participants' daily routines. \add{Users provided their preferred study date and time slots for each part of the day (morning, afternoon, and evening as defined in Figure~\ref{fig:usr-time-dist}). As shown in Figure~\ref{fig:usr-time-dist}, each point represents the start time of a study session, and the respective users were randomly assigned based on their preferences.}

\subsection{Data Capture Apparatus}

We captured acoustic data using the sensing system integrated into \name{} eyeglasses. Additionally, we recorded ground truth video data to annotate the activities. For this purpose, we employed a GoPro HERO9 camera~\cite{gopro} mounted on the participants' chests using a lightweight body mount from the same manufacturer. The camera's horizontal and vertical field of view was set to $118\degree$ and $69\degree$ respectively. It recorded egocentric videos at a resolution of $720$p and a frame rate of $30$ fps. Additionally, participants were provided with Apple AirPods Pro during sessions where they received audio prompts or instructions for specific activities.

\subsection{Study Design}
We conducted a 12-participant \textit{semi-in-the-wild study} followed by an \textit{unconstrained study} with 7 participants. Both of these studies were conducted at participants' homes in unconstrained settings. The design of the study protocols is discussed below.

\subsubsection{User Study Design Considerations}

We conducted two separate studies at participants' homes in a naturalistic setting. Given that the activity set of \name{} encompasses a wide range of activities, obtaining samples of each activity in an uncontrolled setting, where participants are allowed to carry out their daily routines, proved challenging. Therefore, we conducted a semi-in-the-wild study where users were prompted with specific instructions to perform activities, thus covering the tracking set of \name{}. Subsequently, we conducted a second unconstrained study at participants' homes, where they did not receive any particular instructions to perform activities but instead continued with their daily routines.

We utilized a chest-mounted camera to gather reference videos for ground truth annotation. In contrast to other activity recognition systems evaluated in real-world settings~\cite{mydj}, we opted not to rely on self-reporting-based annotation, despite its facilitation of longer duration in-the-wild data collection without significant labeling effort. This decision stems from the impracticality of accurately self-reporting all 27 types of activities throughout a day. Additionally, we aimed to evaluate \name{} based on the inference generated at each second, necessitating fine-grained ground truth annotation from the uncontrolled study, which is only achievable with a body-mounted camera.

\subsubsection{Study - 01: Semi-in-the-wild User Study}

The semi-in-the-wild user study is partitioned into two segments. In the first segment, the participants received audio instructions to perform certain activities by wearing Apple AirPods Pro. The goal of this study segment is to collect data samples of all the activities included in the recognition set of \name{}. Before starting this segment of the study, the participants were briefed about the procedure and familiarized with the audio instructions they were going to receive for each activity.

The activity set was split according to the indoor locations mentioned in Subsec.~\ref{sec:activity-set}. Two sessions of activity data were collected for the bathroom and kitchen locations. The living area activities were divided into two subsets for the convenience of participants, and two sessions of data were collected for each subset. In each session of the data collection process of this segment, the participants received audio instructions for specific activities in random order, and each activity was repeated 5 times within the session. The duration of each repetition of activities spanned from 10 to 30 seconds. The participants were provided minimal instruction regarding the way to perform certain activities so that they could perform their natural body movements. The duration of this segment, comprising a total of eight sessions, is 68 minutes, and each session was 8.5 minutes long. In between each session, the participants were asked to remount the \name{} eyeglasses.

In the second segment of the semi-in-the-wild study, the participants did not receive any prompt or instruction to perform certain activities. They were allowed to perform their regular daily routine. The total duration of this segment was 30 minutes and was divided into three sessions. The participants wore a chest-mounted camera so that the ground truth video could be recorded for activity annotation. When the participants were performing the activities in this segment, the experimenter was present at the participants' home.

\subsubsection{Study - 02: Unconstrained Study at Participants' Home}

We designed a longer-duration unconstrained study with the goal of evaluating the \name{} activity recognition system in the wild for extended hours. Furthermore, the activity set of \name{} (listed in Subsec.~\ref{sec:activity-set}) contains a total of 27 activities, including the \textit{null} class, and it is unlikely to get samples of all those activities in the second segment (30 minutes) of the semi-in-the-wild study in a naturalistic setting. Since the IRB protocol allows a maximum duration of two hours for a single study, we designed this in-the-wild study with a duration of two hours. However, the protocol allows multiple studies with the same participant, and therefore the participants were given an option to take part in multiple consecutive studies. One out of the 7 participants opted for that choice and participated in two consecutive studies which were four hours long in total, and they were compensated twice. Hence, we accumulated 16 hours of in-the-wild data from this study.

We followed the same data collection procedure for the unconstrained study as the second segment of the semi-in-the-wild study. One difference to be noted is that the experimenter left the participants' homes after briefing them and setting up the data collection system. The participants returned the data collection system after two hours. In this unconstrained study, the participants were not instructed with any specifics of the activities to be performed during the study; rather, they were allowed to continue their regular schedule at their homes. Note that we did not permit participants to leave their homes due to legal restrictions on video recording in public places in the country where the study was conducted.

\subsection{Peer-reviewed Data Annotation Protocol}

To provide annotations for active acoustic data via ground-truth egocentric video data, we utilized the ANU-CVML Video Annotation Tool (Vidat)~\cite{zhang2020vidat} to annotate all ground-truth egocentric video data with action annotations. Subsequently, we synchronized the timestamps of the acoustic and video data using a clapping action \add{at the beginning of the recording}, which had a distinct acoustic signature. We then developed separate postprocessing scripts to align video annotations with acoustic echo profiles. To ensure accurate annotation of activities in a naturalistic setting, we implemented a peer-reviewed annotation process. In this procedure, one annotator from the research team labeled the data, while another researcher independently reviewed the annotations and provided feedback. After a phase of revision and approval from the reviewer, the annotated labels were incorporated into the \name{} dataset.
\section{PERFORMANCE EVALUATION}
\label{sec:eval}

We evaluated the performance of \name{} on the data collected in user studies. Our evaluation can be partitioned into two phases. In the first phase, we evaluate the activity recognition performance on the prompted sessions of the semi-in-the-wild user study. Subsequently, we benchmark the performance of \name{} on unconstrained sessions of the user study. For both scenarios, we employed a leave-one-participant-out strategy, to evaluate its performance without the need to collect training data from any new user in new environments.

\subsection{Leave-one-participant-out Evaluation of Prompted Sessions}
\label{subsec:prompt-eval}


We conducted a leave-one-participant-out cross-validation evaluation on prompted sessions of the semi-in-the-wild study. In the initial segment of the semi-in-the-wild study (as described in Sec~\ref{sec:user-study}), involving 12 participants, each participant contributed $8$ sessions. We trained $12$ user-independent models, where, for instance, the model for participant P01 was trained solely on data from P02-P12 and tested on P01, and the same process was repeated for the other participants. The average macro F1-score for each participant ranged from $0.90$ to $0.95$, exhibiting a standard deviation of $0.035$. Across all participants, the average macro F1-score in the leave-one-participant-out evaluation was $0.934$. When examining individual activities, we observed high accuracy (macro F1-score) for all activities, ranging from $0.88$ to $0.97$ across participants in the semi-in-the-wild study with a prediction frequency of 1 Hz. Compared to prior work\cite{lane2015deepear, mollyn2022samosa}, which evaluated activity recognition in a similar manner (albeit with different activities), \name{} has significantly better performance with a wider range of activities. However, we also acknowledge that in this part of the study, participants performed activities following audio stimuli with the presence of a researcher at their home, which may reduce the variance in how they perform activities in real-world settings.




\subsection{Evaluation of User Independent Model on Unconstrained Sessions}
\label{subsec:in-the-wild-res}

    

    

\begin{figure}[h]
    \centering
    \includegraphics[width=0.7\linewidth]{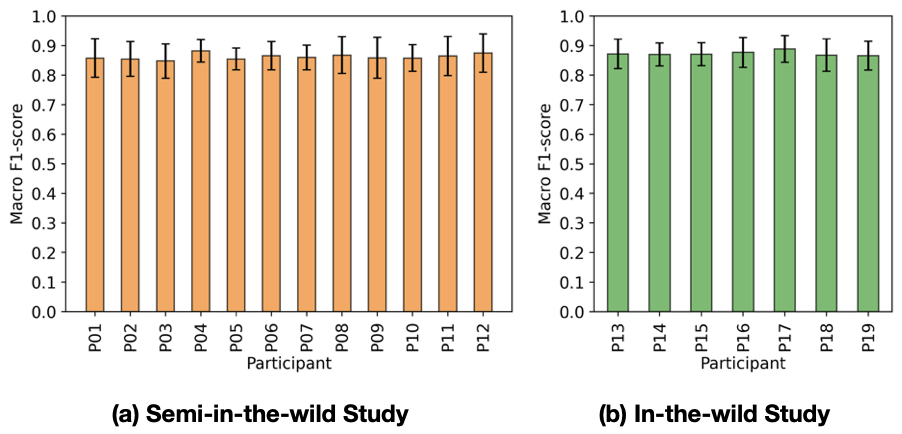}
    \caption{Leave-one-participant-out performance evaluation of \textbf{unconstrained sessions}}
    \label{fig:wild-partwise}
\end{figure}

\begin{figure}[h]
    \centering
    \includegraphics[width=0.6\linewidth]{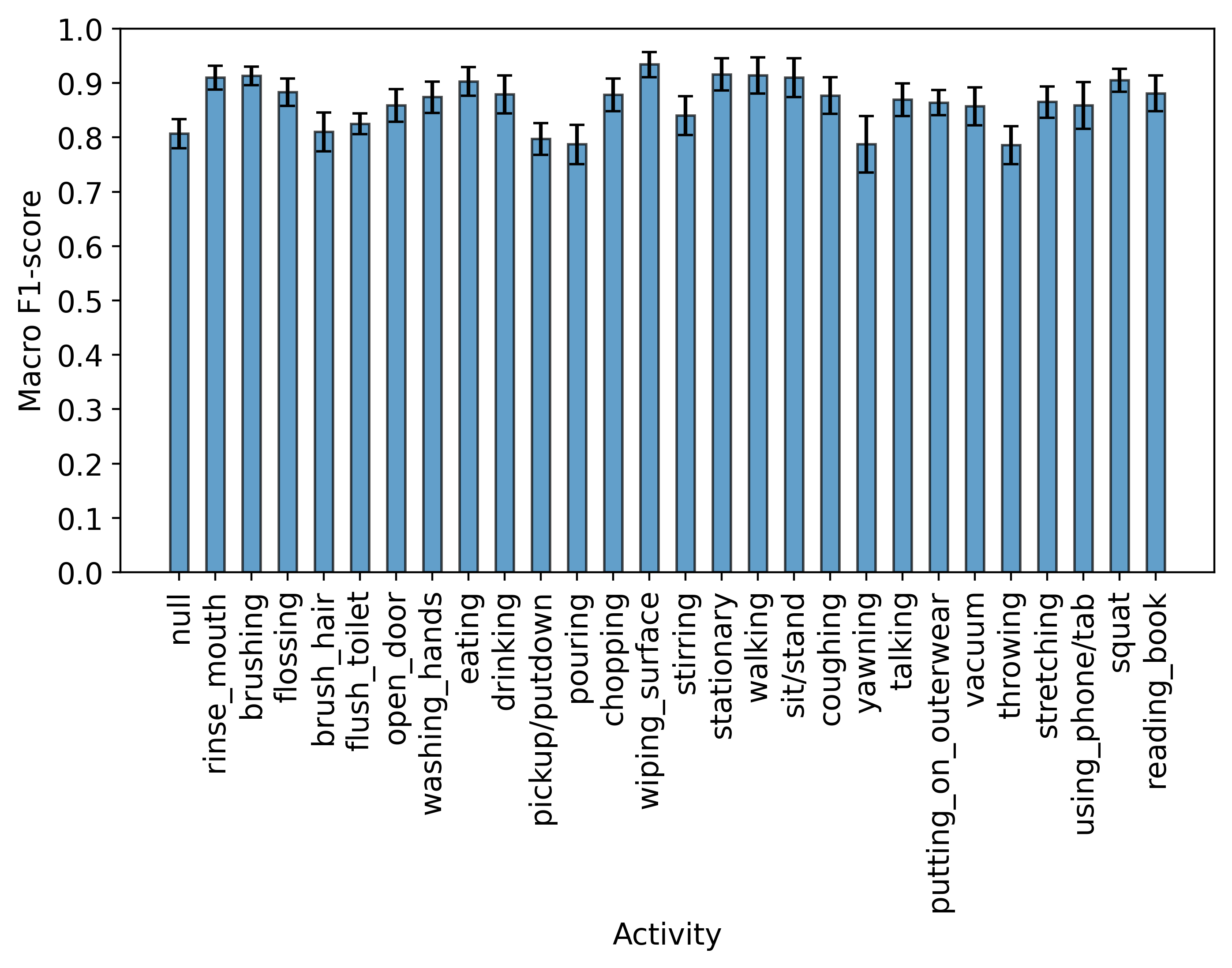}
    \caption{Evaluation of the performance of different activities in \textbf{unconstrained sessions}}
    \label{fig:wild-actwise}
\end{figure}

We further assess our user-independent models through a leave-one-participant-out cross-validation strategy on our dataset of unconstrained sessions. This evaluation aims to gauge \name{}'s performance in real-world naturalistic scenarios. As noted in Sec.~\ref{sec:user-study}, participants continued their regular daily routines at home during the study. Hence, these sessions entail activity samples that might exhibit more diverse motion profiles compared to prompted sessions.

\begin{figure}[h]
    \centering
    \includegraphics[width=0.75\linewidth]{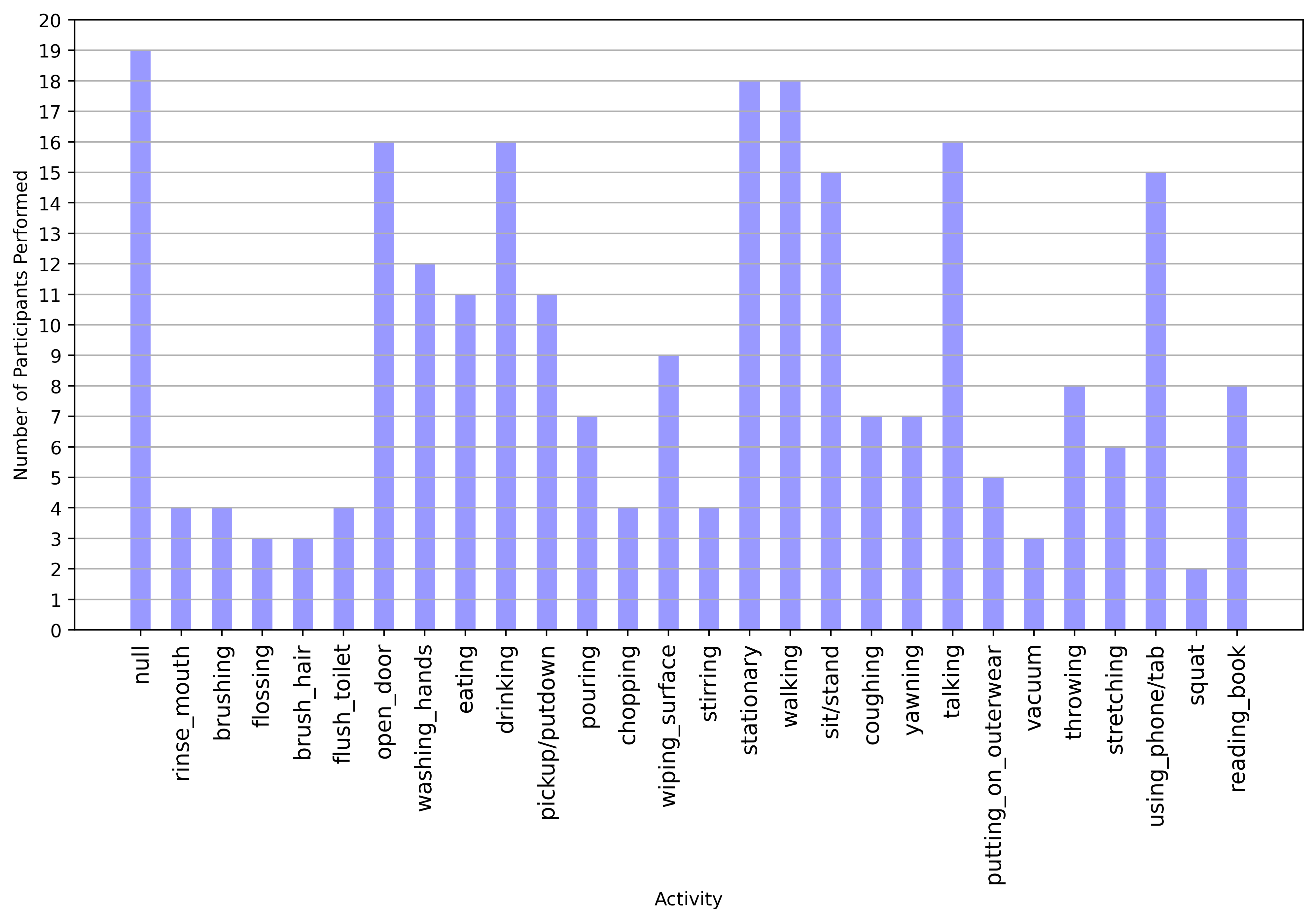}
    \caption{Number of participants per activity during \textbf{unconstrained sessions}: activity labels on the $x$-axis and number of participants on the $y$-axis}
    \label{fig:wild-numact}
\end{figure}

Our evaluation of the unconstrained sessions from both the semi-in-the-wild and the second studies involves two stages. In the initial stage, we take each model trained on prompted sessions and conduct a leave-one-participant-out evaluation (as described in Subsec.~\ref{subsec:prompt-eval}). Subsequently, we fine-tune these models using unconstrained data from the semi-in-the-wild study for other participants before assessing the model on the original participant. For instance, the P01 "prompted" model, trained on P02-P12 "prompted" supervision, undergoes fine-tuning using "unconstrained" supervision from P02-P12, ensuring the exclusion of labels from P01 during model training.

\begin{figure}[h]
    \centering
    \includegraphics[width=1.03\linewidth]{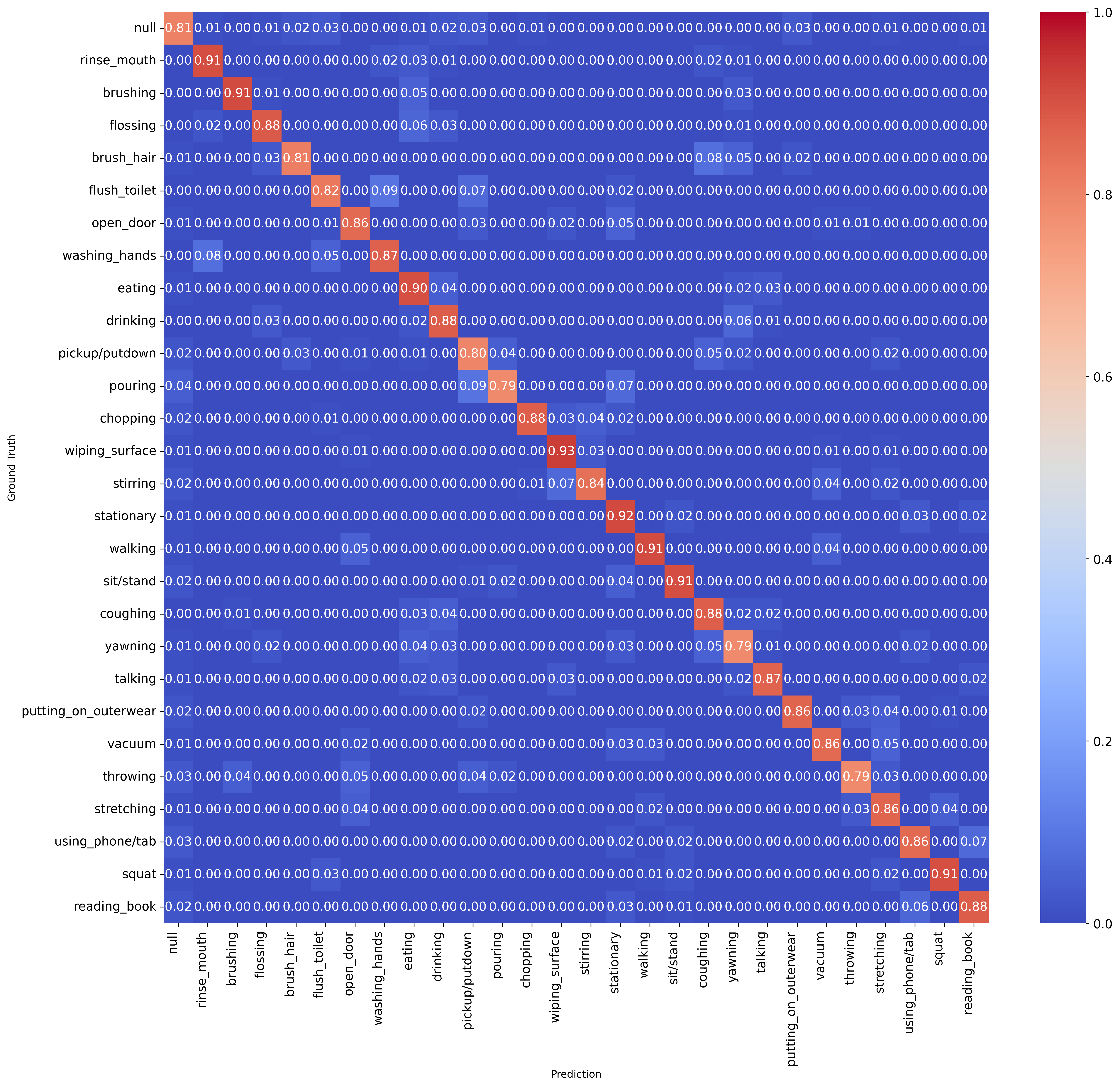}
    \caption{Normalized confusion matrix of leave-one-participant-out user evaluation in \textbf{unconstrained sessions}}
    \label{fig:wild-conf}
\end{figure}

In the second phase of the unconstrained session evaluation, we measure the performance using data from the second study. To assess the unconstrained sessions of users P13 to P19 from the unconstrained study, we first train a fine-tuned model using prompted session data of users P01 to P12 as supervision. Subsequently, we evaluate the model performance using data from individual users of P13-P19 as the test set.

The average macro F1-score and standard deviation for all actions listed in Sec~\ref{sec:activity-set} are reported in Figure~\ref{fig:wild-partwise}. Specifically, Figure~\ref{fig:wild-partwise}(a) and~\ref{fig:wild-partwise}(b) present the average activity recognition performance on semi-in-the-wild and unconstrained study participants respectively. Furthermore, Figure~\ref{fig:wild-actwise} displays the average macro F1-score and standard deviation for each of the unconstrained actions across all participants. Our findings reveal an average F1-score of $0.866$ with a standard deviation of $0.052$ with a predication frequency of 1 Hz. Although the system's performance in an unconstrained scenario is comparatively lower than in prompted sessions, it is apparent in Figure~\ref{fig:wild-conf} that the activity recognition remains accurate, considering the user variability and class imbalance present in the unconstrained sessions. 

The class distribution, indicating the number of participants performing specific activities, is illustrated in Figure~\ref{fig:wild-numact}. The observed accuracy imbalance among different actions within the unconstrained examples likely stems from the natural variability in individual interactions within an open-world setting, influenced by external context. While activities like "rinsing mouth," "brushing," "walking," and "sitting/standing" seem less contextually dependent, actions such as "pouring," "throwing," and "pickup/putdown" might display lower performance due to subtler movements and inherent variability arising from interactions with different objects.

\subsection{Power Signature of the \add{\name{} Sensing System on Eyeglasses Form Factor}}
\name{} system initially consumed $577.8$ mW \add{for collecting data} with the first prototype employing Teensy 4.1. Substituting this microcontroller with a low-power nRF52840 significantly reduced power consumption. We integrated the original speakers and microphones from the initial prototype into the second one, adjusting the gain to ensure identical sound pressure levels (SPL) for both, maintaining emission consistency. The power signature, measured using a CurrentRanger~\cite{current-ranger}, displayed an average operation of $96.5$ mW ($4.0$2 V, $24.0$ mA) while saving all data to the SD card. Furthermore, we conducted long-term stability testing, and the prototype operated continuously for $11.3$ hours using a $290$ mAh $3.7$ V Li-Po battery. This configuration enables a full-day operation on commodity smart glasses or AR glasses. For instance, Google Glass, equipped with a $570$ mAh battery, can support the \name{} sensing system for over $21$ hours if the activity recognition pipeline is the only active process.

\subsection{Latency and System Overhead of \name{} \add{Inference Pipeline} on Mobile Platform}
\begin{table}[ht!]
\centering
\begin{tabular}{@{}lcc@{}}
\toprule
\textbf{}           & \textbf{Non-quantized} & \textbf{Quantized} \\ \midrule
Avg. Macro F1-Score & 0.864                  & 0.772              \\
Size                & 151.1 MB               & 45.4 MB            \\
Inference Time      & 123.1 ms               & 68.4 ms            \\
CPU                 & 14\%                   & 11\%               \\ \bottomrule
\end{tabular}
\caption{\name{} ResNet18 model latency and system overhead on Google Pixel 7 android mobile platform.}
\label{table:latency}
\end{table}

We evaluated the latency and overhead of the \name{} \add{inference pipeline} on the Google Pixel 7 mobile platform. Table~\ref{table:latency} presents the inference time and various parameters of the \name{} ResNet18 model evaluated on an unconstrained dataset with the same protocol described in~\ref{subsec:in-the-wild-res}. Initially, we generated lightweight mobile models from both the original and the 8-bit integer quantized versions of the ResNet18 model using PyTorch Mobile. Subsequently, we conducted inference time benchmarks using a two-second sliding window on a Google Pixel 7 Android phone, performing inference on 1000 samples. The mean inference time is detailed in Table~\ref{table:latency}. The \name{} performance on mobile devices was evaluated by transmitting the acoustic data to a Pixel 7 for inference. The BLE status was set to "connected" since it continuously streams data to mobile devices.

Furthermore, we utilized the Android Profiler~\cite{android-profiler} to evaluate mobile CPU usage and energy consumption. Table~\ref{table:latency} indicates that while the quantized model exhibits lower accuracy compared to its non-quantized counterpart, it demonstrates lower system overhead. This assessment utilized a post-training quantization strategy; employing a quantization-aware training approach for the ResNet18 model might yield improved performance while preserving similar efficiency. The energy consumption was indicated as "light" for both the quantized and non-quantized models by the Android Profiler~\cite{android-profiler}.

\section{ABLATION STUDY}

\subsection{Impact of Sensing Different Body Parts on Activity Recognition Performance}

\begin{figure}[h]
    \centering
    \includegraphics[width=0.85\linewidth]{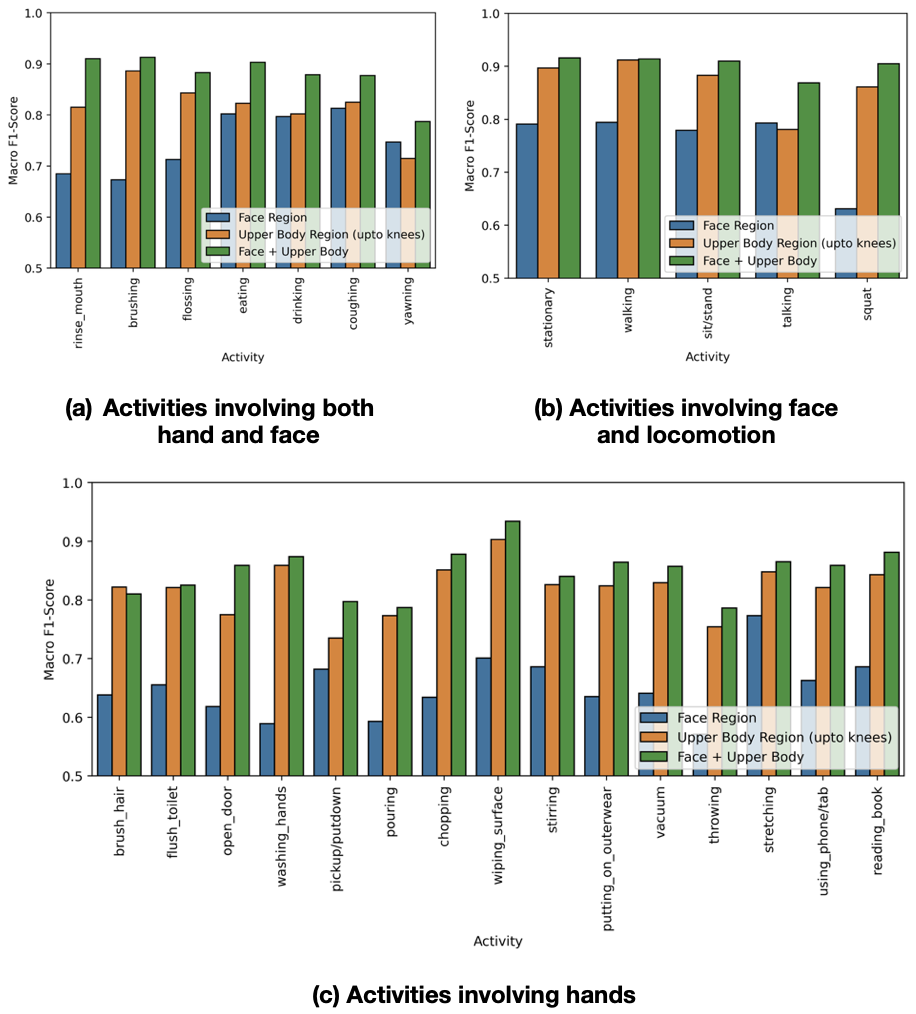}
    \caption{The impact of using acoustic signals corresponding to different body regions on the performance of \name{} is evaluated. The face region comprises the first $50$ pixels in the differential echo profile, covering movement within $17.15$ cm of the sensing system. The upper body region encompasses the remainder of the echo profile sliding window. "Face $+$ Upper Body" denotes the performance of \name{} using the entire sliding window (whose shape is tuned as a hyperparameter).}
    \label{fig:hand-vs-face}
\end{figure}

\name{} relies on tracking the movement of facial and upper body limbs to recognize everyday activities. To assess the impact of different body regions on activity recognition performance, we conduct an evaluation of the \name{} system using acoustic signals solely from the face and upper body regions. We then compare this recognition performance with the evaluation reported in Sec.~\ref{sec:eval}. In order to filter out the acoustic reflection from the face region, we crop the first $50$ pixels from the top to bottom of the $y$-axis of the echo profile sliding window. This $ 50$ pixel $(= 17.15$~cm$)$ approximately represents the face region of the user and the movement from this region is captured in the cropped differential echo profile. In addition, we evaluate the performance of \name{} with the rest of the echo profile sliding window (representing the upper body region up to the knees). We present the performance of \name{} under these scenarios in Figure~\ref{fig:hand-vs-face}.

From the performance reported in Figure~\ref{fig:hand-vs-face}, we observe a sharp degradation in performance if we exclude the movement patterns of the upper body region. Analyzing the activity-wise performance, we note that activities involving obvious facial movements have fewer errors compared to activities that involve upper body movements, such as eating, drinking, talking, etc. On the other hand, we observe less degradation in performance if we exclude the face region movement from the acoustic signal. This observation can be attributed to the fact that most activities in the \name{} dataset involve hand or upper body movement. Overall, based on the evaluation presented in Figure~\ref{fig:hand-vs-face}, we can extrapolate that the combination of reflection patterns from the face and upper body regions yields the best performance for the \name{} system.

\subsection{\add{Impact of Different Placements of the \name{} Sensing System on the Eyeglasses Form Factor}}
\begin{figure}[h]
    \centering
    \includegraphics[width=1.0\linewidth]{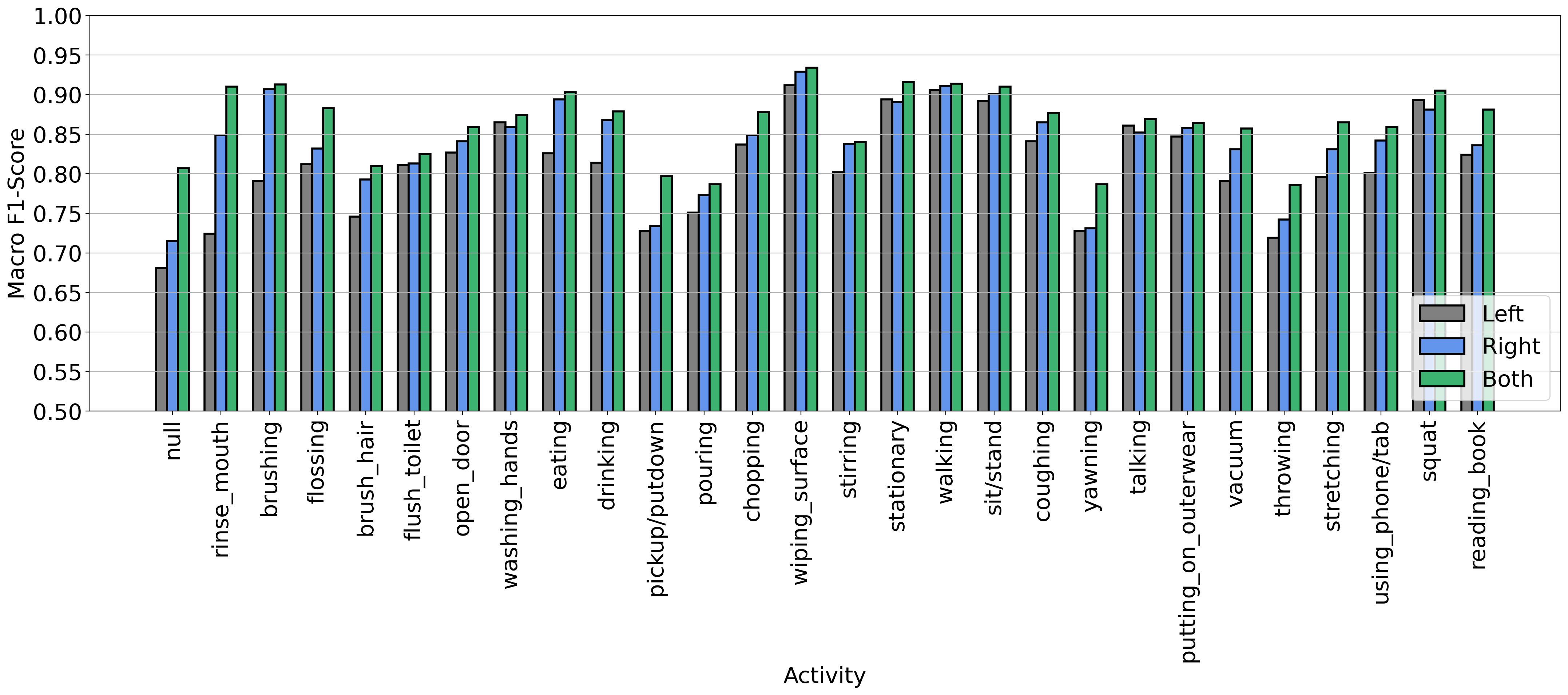}
    \caption{Impact of active acoustic sensing placement on recognition performance of \name{}: \textbf{Left} indicates a speaker (emitting C-FMCW chirps within the frequency range of 18.0-21.0 KHz) and microphone pair on the left hinge of the eyeglasses, \textbf{Right} indicates a speaker (emitting C-FMCW chirps within the frequency range of 21.5-24.5 KHz) and microphone pair on the right hinge of the eyeglasses, and \textbf{Both} indicates both aforementioned speaker-mic pairs are placed on both hinges of the eyeglasses.}
    \label{fig:side-ablation}
\end{figure}

\add{We evaluated the impact of sensing system placement on the performance of \name{} in this experiment. The raw audio data was collected in user studies using the \name{} sensing system placed on both hinges of the eyeglasses. To measure the impact of sensor placement, we utilized the acoustic data from one side, either left or right. We then computed the echo profile and acoustic flow using that data. With only one side activated, we have just one C-FMCW frequency range, yielding echo profiles with one channel containing information on the direct acoustic transmission path (details of transmission paths are mentioned in Sec.~\ref{sec:echo-profile}). We customized the deep learning pipeline to take input sliding windows of the same shape with one channel and evaluated the activity recognition performance of \name{}.}

\add{Figure~\ref{fig:side-ablation} illustrates that placing the sensing system on both hinges of the eyeglasses yields the best performance, with a mean macro F1 score of $0.866 \pm 0.044$ in our leave-one-participant-out evaluation. The right-side placement demonstrated moderate performance, with a mean macro F1 score of $0.838 \pm 0.057$, which is close to the performance of both-side placements. This can be explained by the involvement of the dominant hand in most of the activities in the \name{} recognition set, which was the right hand for all the participants in the study. On the other hand, the left hinge placement demonstrated degraded performance compared to the right one, with a mean macro F1 score of $0.811 \pm 0.062$.}

\subsection{Performance Comparison of Different Deep Learning Encoders}

\begin{table}[h]
\centering
\begin{tabular}{@{}lccc@{}}
\toprule
\textbf{Model Architecture}      & \textbf{\begin{tabular}[c]{@{}c@{}}Number of Parameters \\ (Approx.)\end{tabular}} & \textbf{Prompted Sessions} & \textbf{Unconstrained Sessions} \\ \midrule
MobileNetV2 w/o Self-supervision & 4M                                                                                 & 0.827                     & 0.743                         \\
MobileNetV2 w/ Self-supervision  & 4M                                                                                 & 0.854                     & 0.761                         \\
ConvLSTM                         & 16M                                                                                & 0.879                     & 0.788                         \\
ResNet18 w/o Self-supervision    & 11M                                                                                & 0.912                     & 0.785                         \\
\textbf{ResNet18 w/ Self-supervision}     & \textbf{11M}                                                                                & \textbf{0.934}                     & \textbf{0.866 }                        \\ \bottomrule
\end{tabular}
\caption{Comparison of \name{} performance under different deep learning encoders and training strategies. The number of trainable parameters for each model is reported in millions (M).}
\label{table:models}
\end{table}

We developed a self-supervised deep learning pipeline for \name{}, utilizing ResNet18~\cite{resnet} as the backbone encoder. The performance of this network is compared with architectures having different encoders and training strategies in Table~\ref{table:models}. We present the performance of MobileNetV2~\cite{sandler2018mobilenetv2} and the ConvLSTM architecture (ResNet18 encoder followed by an LSTM decoder with two layers) in Table~\ref{table:models}. We also evaluate the impact of self-supervised pretraining on convolutional encoders (ResNet18 and MobileNetV2). Additionally, the number of trainable parameters (in millions) for each model is reported in Table~\ref{table:models}.

Observing Table~\ref{table:models}, we note that while self-supervision doesn't exhibit significant performance improvement in the controlled sessions, it does demonstrate an impact in maintaining performance in variable unconstrained scenarios. Furthermore, in comparison to the number of parameters of MobileNetV2, ResNet18 has a larger memory footprint. This observation is particularly valuable in the scenarios involving the performance-inference time tradeoff of the \name{} system. Additionally, ConvLSTM exhibits worse performance compared to self-supervised ResNet18 despite having the explicit capacity to model temporal dependency.

\subsection{Comparison of Performance with Prior Systems}

\begin{table}[h]
\scalebox{0.8}{
\begin{tabular}{|l|l|l|c|l|c|l|}
\hline
\textbf{System} & \textbf{\begin{tabular}[c]{@{}l@{}}Sensing\\ Modality\end{tabular}}                        & \textbf{Device Type}                                                                   & \textbf{\begin{tabular}[c]{@{}l@{}}Number of \\ Activities\end{tabular}} & \textbf{Activity Examples}                                                                                                                           & \textbf{\begin{tabular}[c]{@{}l@{}}Performance \\ (Accuracy)\end{tabular}} & \textbf{Study Design}                                                                                                    \\ \hline
BodyScope~\cite{yatani2012bodyscope}       & \begin{tabular}[c]{@{}l@{}}Passive \\ Acoustics\end{tabular}                               & \begin{tabular}[c]{@{}l@{}}Bluetooth \\ headset\end{tabular}                           & 12                                                                       & \begin{tabular}[c]{@{}l@{}}eating, drinking, \\ laughing, coughing\end{tabular}                                                                      & 79.50\%                                                                    & \begin{tabular}[c]{@{}l@{}}Small-scale, \\ In-the-wild, \\ 4 activities\end{tabular}                                  \\ \hline
Ubicoustics~\cite{laput2018ubicoustics}     & \begin{tabular}[c]{@{}l@{}}Passive \\ Acoustics\end{tabular}                               & \begin{tabular}[c]{@{}l@{}}Commodity \\ electronic \\ devices \\ with mic\end{tabular} & 30                                                                       & \begin{tabular}[c]{@{}l@{}}chopping, baby crying, \\ knocking, speech, \\ alarm clock, etc.\end{tabular}                                             & 89.60\%                                                                    & In-the-wild                                                                                                              \\ \hline
PrivacyMic~\cite{iravantchi2021privacymic}      & \begin{tabular}[c]{@{}l@{}}Passive \\ Ultrasonic \\ and Infrasonic \\ sensing\end{tabular} & \begin{tabular}[c]{@{}l@{}}Customized \\ hardware \\ board\end{tabular}                & 10                                                                       & \begin{tabular}[c]{@{}l@{}}mixer, microwave, \\ kitchen sink, shredder, \\ toilet, etc.\end{tabular}                                                  & 95\%                                                                       & \begin{tabular}[c]{@{}l@{}}Homes and \\ commercial \\ buildings\end{tabular}                                             \\ \hline
SAMoSA~\cite{mollyn2022samosa}          & \begin{tabular}[c]{@{}l@{}}IMU and \\ Subsampled\\  Passive \\ Audio\end{tabular}       & Smartwatch                                                                             & 26                                                                       & \begin{tabular}[c]{@{}l@{}}drill, blender, \\ microwave, coughing, \\ toothbrushing, etc.\end{tabular}                                                & 92.20\%                                                                    & \begin{tabular}[c]{@{}l@{}}At participants' \\ home, activities \\ performed \\ according to \\ instruction\end{tabular} \\ \hline
DiffAct~\cite{liu2023diffusion}      & Camera                                                                                     & \begin{tabular}[c]{@{}l@{}}Body-mounted \\ egocentric\end{tabular}                     & 71                                                                       & \begin{tabular}[c]{@{}l@{}}take cup, \\ preparing coffee, etc.\end{tabular}                                                        & 82.20\%                                                                    & Researcher data                                                                                                        \\ \hline
\textbf{\name{}}        & \begin{tabular}[c]{@{}l@{}}\textbf{Active} \\ \textbf{Acoustic} \\ \textbf{Sensing}\end{tabular}                      & \begin{tabular}[c]{@{}l@{}}\textbf{Commodity} \\ \textbf{Eyeglasses}\end{tabular}                        & \textbf{27}                                                                       & \begin{tabular}[c]{@{}l@{}}\textbf{toothbrushing,} \\ \textbf{flossing, eating,} \\ \textbf{drinking, washing} \\ \textbf{hands, coughing,} \\ \textbf{reading book,} \\ \textbf{wiping surface, etc.}\end{tabular} & \textbf{93.40\%}                                                                    & \begin{tabular}[c]{@{}l@{}}\textbf{Semi-in-the-wild,} \\ \textbf{at participants' home,} \\ \textbf{naturalistic setting}\end{tabular}               \\ \hline
\end{tabular}
}
\caption{Comparison of performance of activity recognition systems with similar sensing modality. Note that the systems were not evaluated on the same dataset. Therefore, numerical differences may not provide a fair comparison. User study evaluation information is also provided in the table.}
\label{table:comparison}
\end{table}

We compare \name{} with other activity recognition systems in Table~\ref{table:comparison}. \name{}, an active acoustic sensing-based activity recognition system, is compared with systems utilizing audible acoustic data, passively sensed inaudible (ultrasonic and infrasonic) acoustic data, motion data (captured through IMU), and egocentric camera data.

Activity recognition systems rely on passively sensed audible signals (such as BodyScope~\cite{yatani2012bodyscope}, Ubicoustics~\cite{laput2018ubicoustics}) or inaudible acoustics (PrivacyMic~\cite{iravantchi2021privacymic}). These systems operate on the premise that different actions generate distinct acoustic signatures in the environment. Consequently, their activity sets include actions like chopping, baby crying, or using a mixer. However, many daily activities, such as wearing clothes, reading, or flossing, do not produce discernible sounds suitable for activity inference. Moreover, these activities may not necessarily involve the user wearing the device, especially with wearables. Additionally, previous studies have mainly focused on smaller-scale experiments with specialized sensors or large-scale video-only benchmarks (e.g., DiffAct~\cite{liu2023diffusion}), with few considering real-world settings.

On the other hand, \name{} can track fine-grained actions such as flossing or wiping surfaces since its sensing mechanism relies on tracking the movement of different body parts simultaneously. It is evident from Table~\ref{table:comparison} that \name{} achieves higher recognition performance (Macro F1-score over $90\%$) in naturalistic environments, with an extensive set of fine-grained everyday activities. \name{}'s utilization of an active acoustic sensing mechanism enables the capture of signals representing fine-grained movements that passive sensing-based systems may miss. For instance, it can recognize actions occurring outside the frame of an egocentric camera or actions with minimal audio cues, which passive acoustic sensing may struggle to detect accurately. Additionally, passive acoustic sensing-based methods are vulnerable to changes in environmental parameters, hindering their ability to generalize signals across different environments and making deployment in the wild challenging.


\section{DISCUSSION}
\label{sec:discussion}

\subsection{Visual Explanation of Learned Parameters of \name{} Model through Saliency Analysis}
\label{sec:gradcam}
    
\begin{figure}[h!]
    \centering
     \includegraphics[width=0.85\linewidth]{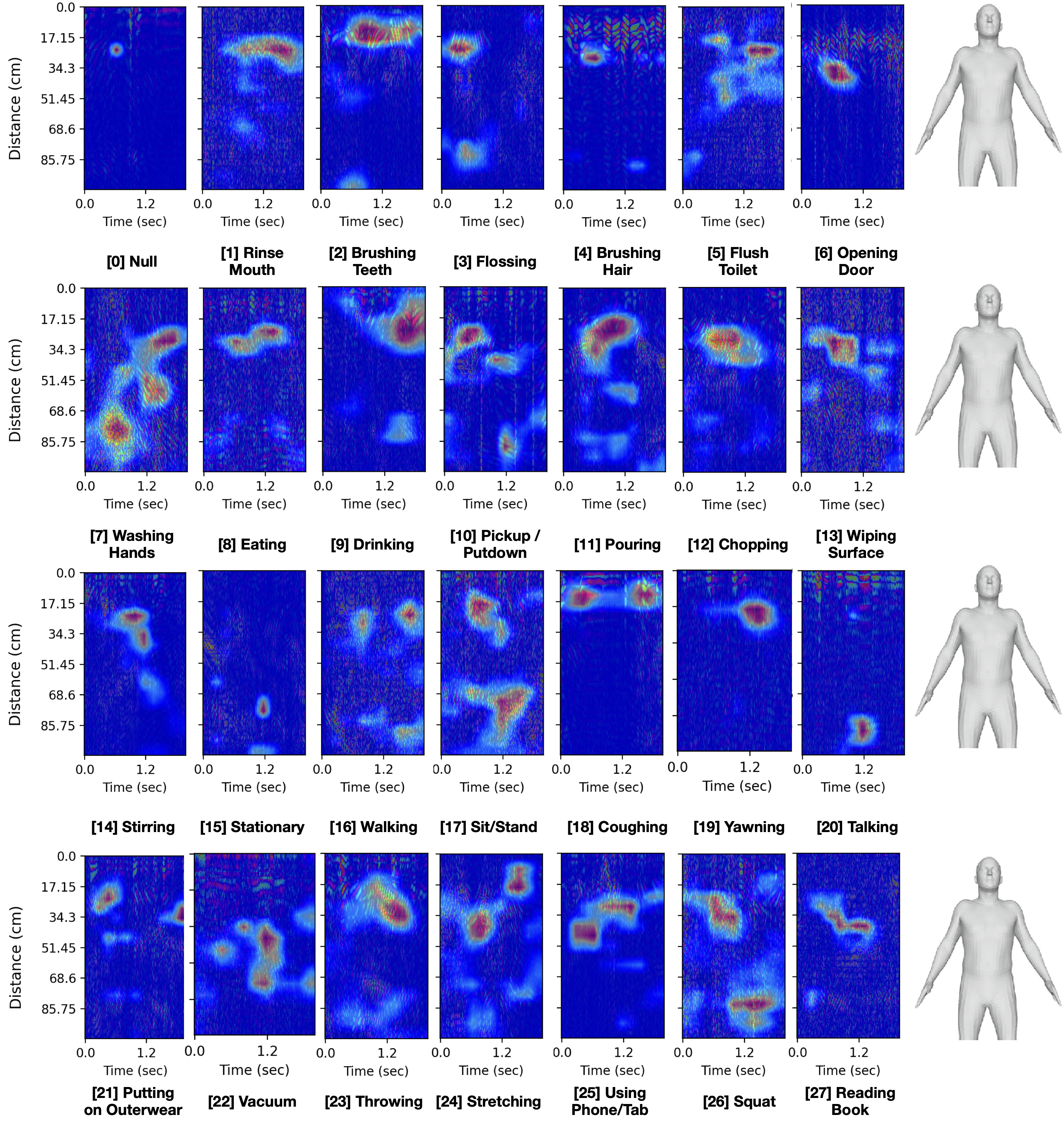}
     \caption{Selected Grad-CAM~\cite{grad-cam} heatmaps overlaid on differential echo profiles. As we have a 4-channel input and Grad-CAM aggregates heat maps by channel, we overlay the same (smoothed) heatmap across all four channels. Redder values from the smooth interpolation correspond to higher significance towards class prediction, while bluer values indicate lower significance. We display a SMPL~\cite{smpl} mesh of the body region covered by the sensing range of \name{}'s sliding window input in the rightmost column.}
     \label{fig:gradcam}
 \end{figure}

Our user study demonstrated competitive accuracy in recognizing 27 everyday activities by learning patterns from reflected inaudible acoustic waves around the body. The reflected acoustic signals are formed with complex multipath echoes from both the body and the surrounding environments. Additionally, the environment can occasionally generate acoustic signals with frequency components that overlap with our emitted frequency range. Therefore, received acoustic signals contain rich information about the environment in addition to the body poses. 

A common critique of many data-driven sensing systems is that machine learning algorithms operate as a black box, making it difficult to understand what and how they learn. Given the complex reflections forming the received acoustic signals, it is critically important to understand what the deep learning algorithm actually learns from these signals.

To address this, we conducted saliency analysis using Gradient-weighted Class Activation Mapping (Grad-CAM)~\cite{grad-cam} to visualize the ResNet18 encoder's final convolutional layer. Overlaying feature heatmaps on acoustic flow or differential echo profile channels confirms the model's ability to capture the movements of body parts and areas related to each activity. Figure~\ref{fig:gradcam} demonstrates Grad-CAM~\cite{grad-cam} for all activities in the dataset.

As shown in Figure~\ref{fig:gradcam}, various activities activate distinct regions within the echo profile sliding window in the neural networks. The $y$-axis represents the distance from reflected echoes to the sensor, while the $x$-axis denotes the timeline. To help comprehension, we aligned the $y$-axis with a human body model. Colors in the figure indicate body movement intensity, with red indicating high intensity and dark blue indicating lower intensity.

The visualization clearly shows that the deep learning algorithms allocate significantly higher attention to areas with heavy body movements (e.g., face, chest, arms) during different activities. For instance, in the activity of brushing teeth, repetitive movement near the face region is evident, with the ResNet18 encoder displaying higher gradient values (indicative of heightened attention) in that area around the face ($5$ cm-$17.15$ cm) to infer the activity. 

While some activities share similar arm movement patterns, fine details in reflected acoustic waves around the face differentiate them. For instance, both drinking and eating involves moving the arm toward the mouth. However, after feeding during eating, the hand usually moves downward, leaving only mouth movements, visible as subtle highlights under $10$ cm. Conversely, during drinking, the hand and container (e.g., cup) remain around the mouth, evident as a large area of heated sections ($15$-$30$ cm). Additionally, the activity of opening a door demonstrates hand movement to unlock the door, followed by movement to enter the room.

This visualization confirms that our self-supervised model focuses on different explainable regions of the echo profile within the input sliding window to infer everyday activities. It underscores the importance of capturing detailed movements in both the upper body and face areas to distinguish fine-grained activities such as eating versus drinking, which may not be distinguished in isolation.

\subsection{Deploying \name{} on Other Wearable Form Factors}
\name{} employs an active acoustic sensing system integrated into eyeglasses to emit inaudible acoustic signals around the body and capture the reflected waves. Due to the low-power and compact nature of the microphones and speakers, this technology can be easily adapted to other wearable form factors for tracking human activities using a similar sensing principle. Here, we evaluate the feasibility of implementing \name{} on different form factors. Two critical considerations for applying \name{} to wearable form factors are: 1) the ability to continuously emit acoustic waves around both sides of the upper body and head, and 2) the stability of the wearable form factor during various activities.

Based on these criteria, we anticipate that \name{} can be seamlessly deployed with minimal modifications on wearable form factors worn around the head (pending training data collection for each form factor), such as necklaces, earbuds, or the recently introduced Humane AI Pin~\cite{ai-pin} worn on the chest. These form factors provide similar stability to eyeglasses while continuously covering both the upper body and head to track detailed facial and upper body movements.

Wearable form factors attached to limbs, such as wristbands, armbands, or rings, pose greater challenges for applying \name{} for activity recognition with a single device. Limbs exhibit a wide range of movements, complicating signal interpretation as the source moves concurrently with other body parts. Additionally, signals emitted from one side of the limb may be obstructed by the body, hindering sensing on both sides. Moreover, capturing facial movements from limb-mounted devices is challenging due to the relatively long distance and potential blockages. These issues may be mitigated with updated algorithms and hardware designs, which we plan to explore in future research.

\subsection{Model Quantization and MCU Inference}
To explore the feasibility of deploying our framework on glasses, we implemented the model pipeline on the MAX78002 microcontroller unit (MCU), leveraging its built-in ultra-low-power CNN accelerator. Initially, we quantized the model by converting high-precision floating-point model parameters to 8-bit integers, a necessary step for deployment on the MAX78002 MCU. Subsequently, we generated a C program for the quantized model inference using the ai8x~\cite{ai8x-repo} library provided by the MCU manufacturer. Due to hardware constraints, certain adjustments were made to the model pipeline for compatibility. Notably, 2D convolution kernel sizes were limited to $(1 \times 1)$ or $(3 \times 3)$, with fixed stride size at $(1 \times 1)$. Additionally, the fully connected layer was capped at a maximum of $1024$ input neurons on the chip. 

\add{Although we successfully ran the \name{} model on the MAX78002 MCU, the computation of the echo profile and inference of one sliding window through the quantized ResNet18 model on the MCU resulted in a mean latency of 1085.54 milliseconds or 1.09 seconds, yielding a refresh rate of 0.92 Hz. Since we predict activity labels for each 2-second sliding window with 50\% overlap, the resulting refresh rate is slower than what is required for real-time inference. The major bottleneck in this process is the computation of the echo profile from the received signal, as this process is not a neural network operation supported by the MAX78002 MCU accelerator. This issue can be alleviated by transforming the cross-correlation operation into an MCU-compatible and accelerated convolution operation, which we leave for future work.}

\subsection{Robustness to Ambient Noise}
The \name{} system utilizes active acoustic sensing to detect daily activities by monitoring body motions. We evaluated its resilience to environmental noise across $19$ participants' homes during the studies. Despite varying environmental factors such as HVAC, running water, TV, and ambient sounds (as detailed in Sec.~\ref{sec:user-study}), \name{} maintains consistent performance independent of environmental settings. This resilience stems from its reliance on ultrasonic frequencies ($18$ kHz to $24.5$ kHz), which exceed most environmental noise sources (recorded at frequencies below $7.5$ kHz).

\add{To confirm this hypothesis, we conducted a noise injection experiment. In this experiment, we overlaid noise from five sources commonly found in households onto the acoustic data recorded with \name{}'s sensing module. These noise sources are people talking ($61.1$ dB), TV/radio ($68.8$ dB), music ($71.5$ dB), pets (dogs barking at $82.1$ dB and cats meowing at $51.0$ dB), and kitchen appliances (microwave at $60.5$ dB and blender at $88.7$ dB). We then computed the echo profile (as described in Sec.~\ref{sec:echo-profile}) and fed it into the \name{} deep learning framework to evaluate the impact of environmental noise.}

\add{To evaluate the impact of different noises on the results, we conducted a one-way repeated measures ANOVA. The $F$-ratio was computed as $0.526$, with degrees of freedom between groups being $5$ and degrees of freedom within groups being $108$. The $p$-value was measured as $0.13 > 0.05$, indicating no statistically significant difference between raw and noise-injected acoustic data for tracking everyday activities using the \name{} system. This lack of significant difference can be attributed to the bandpass filter in the \name{} pipeline, which cancels out the audible frequency range where the injected noises occur. Nevertheless, further investigation into the impact of ambient noise is crucial to evaluate performance under real-world usage scenarios, which we leave for future work.}

\subsection{\add{Effectiveness under Multi-user Scenario}}
\add{Although \name{} is a wearable system designed to track the everyday activities of the person wearing the eyeglasses, the system can be influenced by the movements of other people and objects surrounding the user, as it relies on the reflected acoustic signals of the transmitted wave. The input to the \name{} deep learning pipeline is restricted to a sensing range of 101.185 cm, and theoretically, it should not contain any information from reflected signals beyond this range.}

\add{In the 19-person user study conducted at participants' homes to evaluate the efficacy of the \name{} activity recognition framework, we did not impose any restrictions on the number of other people who could be present at home during the study. To further evaluate the multi-user scenario, we conducted a study with 3 participants from the research team. In this evaluation, each person wore the device and performed 21 activities from the \name{} tracking set (excluding bathroom activities, detailed in Sec.~\ref{sec:activity-set}). While one person performed the activities, the other two were present in the same room, both moving around and staying in place randomly. Additionally, a pedestal fan with a diameter of 16 inches and a height of 47 inches was randomly turned on and off in the same room.}

\add{Our leave-one-participant-out evaluation on these 21 activities yielded a mean macro F1 score of $0.857 \pm 0.029$ across the three users, which is close to the accuracy of the \name{} system evaluated in the 19-person user study. However, we did not evaluate the scenario where multiple users in the same room each wore a \name{} prototype. We leave this evaluation for future investigation.}

\subsection{Health Implication \add{and Usability}}
\name{} emits FMCW-encoded ultrasonic waves for active acoustic sensing. To assess health implications, we measured the transmitted signal intensity using a CDC-provided mobile app~\cite{cdc-app}. The resultant intensity is $68$ dB(A), well below the $85$ dB limit set by NIOSH~\cite{murphy2002revisiting}. Research~\cite{moyano2022possible} on MHz-range ultrasonic exposure suggests muscle tissue discomfort. However, \name{} operates in the KHz range just above the audible threshold, with no reported issues in this range. 

\add{A major concern of active acoustic sensing systems is frequency leakage into the audible range, potentially creating noise from the transmitter. To evaluate this, we collected feedback from participants through a questionnaire after the study. All 19 participants reported that they did not hear any noise or audible sound from the sensing system while wearing the \name{} form factor. Additionally, none reported any comfort issues. However, two participants mentioned that the glass frame size did not fit their faces, four reported needing their prescription lenses while performing the activities, and one reported that the chest-mounted camera interrupted activities such as eating and reading.}

\add{Future investigation should focus on building an attachable sensing module that can be used with any glass frame and integrating the ground truth collection module as an egocentric camera into the eyeglasses form factor. Future studies will also explore potential audibility among animals and children despite its inaudibility to adults for long-term usage.}



\subsection{Privacy Preservation}
\name{} utilizes ultrasonic frequency range ($18$ KHz to $24.5$ KHz) to transmit and receive signal. As mentioned in the description of the sensing system in Sec.~\ref{sec:sensing-mechanism}, we apply a bandpass filter on the audio received by the microphone to ensure that \name{} does not access the audible frequency range to infer activities. Since \name{} does not require any passively sensed audible acoustic signal, the system does not compromise user privacy by processing sensitive conversation information. Furthermore, the potential of adopting a customized ultrasonic speaker and microphone can further remove the possibility of collecting audible sound.

\subsection{Potential Real-world Application}

The promising performance of \name{} recognizing 27 activities in the wild using low-power and minimally obtrusive glasses will significantly lower the barriers to logging everyday activities. It would further create opportunities for many downstream applications that are based on tracking one or multiple types of activities. Here we list a few sample applications :  

\subsubsection{\add{High-resolution Behavior Data in the Wild for Health Monitoring}}

\add{\name{} can be used to journal various everyday activities for different purposes. For instance, journaling food intake behavior is crucial in combating eating disorders, often recorded manually. Previous systems required multiple sensors, had low time resolution (e.g., recognizing a meal every 10 minutes)~\cite{mydj}, or needed training data from a user. In contrast, \name{} can recognize eating moments at 1 Hz with over 90\% F1 score in real-world settings without needing training data from a new user, facilitating immediate deployment for eating journaling practices.}

\add{Additionally, \name{} can track 27 everyday activities, many related to health behaviors. Automatically logging these activities can help researchers and clinicians better understand a user's activities in the wild. For instance, eating and drinking behavior can be analyzed for eating disorders and hydration levels, while activities like brushing teeth, flossing, and rinsing the mouth can be tracked with low error rates, aiding dentists in monitoring patient dental behavior.}


\subsubsection{Tracking Other Activities}
In our study, we were able to track 27 activities. However, our system has the potential to recognize other activities that involve movements on the upper body and face. Researchers can potentially replicate our system and customize the frameworks to detect the activity of their interest. For instance, this system can be easily used to automatically track and log the duration and types of the user's exercise routines.

\subsection{Limitations and Future Work}
\label{sec:limitation}
Our method is currently limited in the following ways, which we aim to further explore in the future: 

    \subsubsection{Scope of Activity Information in the Dataset}
    \name{} recognizes 27 distinct everyday activities within its dataset. However, certain activities in the dataset exhibit variability in execution. For example, actions like yawning or pouring can vary based on contextual factors, which can affect system performance in real-world settings. Addressing this diversity might benefit from a larger dataset and a foundational deep learning feature extractor. Moreover, the activity recognition pipeline of \name{} lacks external contextual information. Integrating GPS or motion data for fine-grained head or hand movements could aid in understanding environmental affordances and improving activity detection.
    
    \subsubsection{Usage of Differential Echo Profile Only}
    Our system relies solely on the differential echo profile, which may miss static activities with consistent poses. While incorporating the original echo profile might address this, our pilot studies revealed reduced performance and user-dependent features, whereas the differential profile remained more user-independent.
    
    \subsubsection{Multi-label or Concurrent Activity Detection}
    Real-world scenarios involve concurrent activities, a challenge yet to be explored in wearable technology. We aim to explore multi-label classifiers leveraging our system's superior performance.
    
    \subsubsection{Reducing Classification Error during Transitions}
    Many of the misclassification errors occurred when the participants were in transition between two activities, as our system makes predictions every second. In the future, these errors can be easily optimized by developing a state machine, or as simple as a majority-vote mechanism. 

\section{CONCLUSION}
This paper introduces \name{}, a low-power and unobtrusive action recognition system that employs acoustic sensing on smart glasses. Extensive experiments involving $19$ participants in real-world settings showcase \name{}'s adeptness in distinguishing a diverse range of everyday actions across different environments, \add{achieving a mean accuracy of 86.6\% in unconstrained user-independent evaluations}. We envision \name{} as a straightforward and efficient supplementary modality for egocentric action recognition, addressing concerns regarding privacy.

\section*{ACKNOWLEDGMENTS}
This project was supported by the National Science Foundation Grant No. 2239569 and partially by the Cornell University IGNITE Innovation Acceleration Program. ChatGPT was used to polish the writing of the paper.

\bibliographystyle{ACM-Reference-Format}
\bibliography{main}

\end{document}